\pdfoutput=1
\documentclass[twocolumn,trackchanges]{aastex63}

\usepackage{amsmath}

\newcommand{\ie}{{\rm i.e.~}}

\submitjournal{ApJ}
\accepted{June 9, 2020}

\shorttitle{Lensed GRBs}
\shortauthors{Ahlgren et al.}
\graphicspath{{./}{figures/}}
\begin{document}

\title{A search for lensed gamma-ray bursts in 11 years of observations by {\it Fermi} GBM}

\correspondingauthor{Bj\"orn Ahlgren}
\email{bjornah@kth.se}

\author[0000-0003-4000-8341]{Bj\"orn Ahlgren}
\affil{KTH Royal Institute of Technology, Department of Physics,\\ 
and the Oscar Klein Centre \\
AlbaNova, SE-106 91 Stockholm, Sweden}
%\collaboration{(Fermi collaboration)}

\author[0000-0003-0065-2933]{Josefin Larsson}
\affil{KTH Royal Institute of Technology, Department of Physics,\\ 
and the Oscar Klein Centre \\
AlbaNova, SE-106 91 Stockholm, Sweden}

\begin{abstract}
Macrolensing of gamma-ray bursts (GRBs) is expected to manifest as a GRB recurring with the same light curve and spectrum as a previous one, but with a different flux and a slightly offset position. Identifying such lensed GRBs may give important information about the lenses, cosmology, as well as the GRBs themselves. 
Here we present a search for lensed GRBs among $\sim 2700$ GRBs observed by the {\it Fermi} Gamma-ray Burst Monitor (GBM) during 11 years of operations. 
To identify lensed GRBs, we perform initial cuts on position, time-averaged spectral properties and relative duration. We then use the cross-correlation function to assess the similarity of light curves, and finally analyze the time-resolved spectra of the most promising candidates. 
We find no convincing lens candidates. The most similar pairs are single-pulsed GRBs with relatively few time bins for the spectral analysis. This is best explained by similarities within the GRB population rather than lensing. 
However, the null result does not rule out the presence of macrolensed GRBs in the sample. In particular, we find that observational uncertainties and Poisson fluctuations can lead to significant differences within a pair of lensed GRBs. 
\end{abstract}


\section{Introduction} \label{sec:intro}
The possibility of lensing of GRBs was originally discussed in the context of testing whether GRBs are located at cosmological distances \citep{Paczynski1986,Mao1992}. With the cosmological origin firmly established, we are now approaching sample sizes where the probability of observing a lensed GRB is no longer negligible.  In this paper we focus on the case of macrolensing, i.e.~strong lensing producing image separations of order arcseconds \citep{Treu2010}.  One of the motivations for searching for lensed GRBs is that the magnification effect means that GRBs located at high redshift can be studied in detail. Depending on the properties and data quality of any lensed GRBs identified, they may also provide information about dark matter distributions and constraints on the Hubble constant  (e.g., \citealt{Oguri2019}). 

Macrolensing of a GRB is expected to manifest as a GRB recurring with the same light curve and spectrum as a previous GRB, but with a different flux and a small positional offset. The time interval between the GRBs may range from days to years.  A challenge in identifying such lenses is that the angular separations between the GRB images are typically smaller than the localization uncertainties of current gamma-ray detectors, which range from arcminutes for {\it Swift}~BAT (and improves to arcseconds when subsequently detected by XRT, \citealt{Burrows2005}),  to several degrees or more for  {\it Fermi}~GBM \citep{Goldstein2019}.  While lenses with large separations could in principle be resolved with {\it Swift},  the relatively small field of view of BAT means that the probability of detecting a lensed pair within this sample is low \citep{Li2014}. 

A more promising approach to identify lensed GRBs is to rely on the assumption that every GRB has a unique light curve and spectral evolution.  An interesting precedent for this is set by the gravitationally lensed blazar B0218+357, which displays double images separated by $\sim 0\farcs{3}$ \citep{Odea1992}. While  {\it Fermi}~LAT does not resolve the images, gamma-ray flares have been observed to repeat with a time-delay of 11.5 days in the combined light curve \citep{Cheung2014}. The main differences between this example and the expectations for lensed GRBs are a longer duration of the emission and a larger emission region for the blazar, with the latter having implications for the effects of so-called microlensing, discussed below in the case of GRBs.

Previous searches for macrolensed GRBs have all yielded null results. This includes searching $\sim 2100$ GRBs observed by {\it BATSE} \citep{Li2014}, $\sim 2300$  GRBs observed by {\it Konus-Wind} \citep{Hurley:2019km}, as well as two smaller samples of GRBs observed in the first years of {\it Fermi}~GBM \citep{Veres2009,2011AIPC.1358...17D}.  These searches were all based on comparisons of light curves and time-averaged spectra of GRBs with overlapping positional uncertainty regions. In this paper we present a search for lensed GRBs among $\sim 2700$ GRBs observed by {\it Fermi}~GBM during 11 years of operations. The fourteen detectors that comprise the {\it Fermi}~GBM continuously observe the entire unobscured sky and offer a high sensitivity over the 8~keV -- 40~MeV energy range \citep{2009ApJ...702..791M}, providing good conditions for identifying lenses. We also extend the methodology compared to previous works by considering time-resolved spectra and assessing the similarities of the most promising candidates using simulations.

 It is possible that one or both GRBs in a lens pair are affected by additional lensing by smaller objects in the lens galaxy, the impact of which depends on the nature of those objects. Microlensing by stars leads to a smearing of light curves on millisecond times scales \citep{Williams1997}, while so-called millilensing by compact globular clusters or massive black holes ($M\gtrsim 10^6~{\rm M_{\odot}}$) leads to emission episodes within a GRB repeating with a time delay longer than seconds \citep{Nemiroff2001}. Lensing by intermediate mass black holes would lead to smearing/echoes on time scales between these extremes \citep{Ji2018}. Confirming that differences between a pair of lensed GRBs are due to any type of small-scale lensing would require redshift information, which is only available for $\sim 5\%$ of the sample. We therefore do not consider these effects in our search, although we note that lens candidates are unlikely to be rejected due to microlensing given our choice of light curve binning. In summary, we confine ourselves to considering macrolensing, while making minimal assumptions about the lenses. This means that we consider well-separated light curves and that we set no a priori upper limit to the possible time delays. Throughout this paper, we will use ``lensing" to refer to this kind of macrolensing.

Below we first describe the data used for the analysis in Section~\ref{sec:data}, describe the selection of lens candidates in Section~\ref{sec:methods} and analyze the final candidates in Section~\ref{sec:FinalLensCands}.  We present and discuss the results in Section~\ref{sec:Discussion}, and provide a summary in Section~\ref{sec:summary}.

\section{Data} \label{sec:data}
We base the search for lens candidates on data from \textit{Fermi} GBM. Specifically, we use time-tagged event (TTE) data, which contains the arrival times of individual photons with a precision of 2 $\mu$s, as well as information regarding in which of the 128 energy channels the photon registered. When performing time series analysis, we consider data from the NaI detectors, whereas we use data from both the NaI and BGO detectors for the spectral analysis, as described in Sections~\ref{sec:lightcurves} and~\ref{sec:timeResolvedSpectra}, respectively. We obtain information about the localizations from an online compilation\footnote{\url{https://icecube.wisc.edu/~grbweb_public/index.html}} that also includes localizations by other telescopes than \textit{Fermi}.

We downloaded all GRBs detected before 2020-01-09 from the online \textit{Fermi} GBM catalog using \textit{3ML} \citep{2015arXiv150708343V}. The resulting sample contains 2712 GRBs, which corresponds to over 3.6 million unique pairs. Although it is in principle possible to also include data from other telescopes, this would complicate the analysis significantly, since different instruments operate in different energy intervals and at different efficiencies.

\section{Selection of lens candidates} \label{sec:methods}
In this Section we describe the methods used to search for pairs of GRBs that are consistent with a gravitational lensing scenario and a common physical origin. As mentioned in Section~\ref{sec:intro}, macrolensing will yield well-separated identical light curves with identical spectra. However, in practice, we do not expect the light curves or spectra to be identical due to observational uncertainties. These include the Poisson nature of the detector counts, the angles between the detectors and the source, varying backgrounds, as well as different flux levels due to the lensing. Additionally, most GRBs observed by GBM have poor localization and no redshift measurements.
We have adopted our methods to take these observational uncertainties into consideration. 
We begin by making cuts to the sample to remove burst pairs that are clearly not consistent with the lensing scenario (\ref{sec:samples}). We proceed by comparing the light curves (\ref{sec:lightcurves}) and finally consider the time-resolved spectra (\ref{sec:timeResolvedSpectra}) of each GRB pair.

\subsection{Initial cuts} \label{sec:samples}
We consider the position, relative duration, and spectral information of each burst pair to remove obviously non-lensed pairs from the sample. Since a full analysis of 3.6 million pairs can become computationally expensive for parts of the analysis, we only perform the subsequent analysis on pairs that passed the previous cuts. In addition to the sample consisting of lens candidates, we also construct a reference sample of GRB pairs that we know are not lens pairs. We refer to the samples as the lens-candidate sample (L) and the non-lensed sample (NL). 

\textbf{Position:} 
This is the only variable we can use to completely rule out a lensing scenario. In order to make a cut based on position, we attribute a circular uncertainty region to each burst based on the statistical and systematic uncertainty of the localization. We set the radius of the combined uncertainty region to $\sigma_{\text{tot}} = \sqrt{(2\sigma_{\text{stat}})^2 + \sigma_{\text{syst}}^2}$, where $\sigma_{\text{stat}}$ is the 68\% confidence level statistical uncertainty and $\sigma_{\text{syst}}$ is the systematic uncertainty. The original localization algorithm of GBM had large systematic uncertainties (up to $14^{\circ}$, \citealt{Connaughton:2015gp}) and we therefore conservatively set $\sigma_{\text{syst}} = 14^{\circ}$. For GRBs localized by other instruments (mainly the {\it Neil Gehrels Swift Observatory}), we set $\sigma_{\text{syst}} = 0^{\circ}$, since the systematic uncertainties are comparatively small. Although the GBM localization has been improved, both by the BALROG algorithm \citep{Burgess:2018get} and the updated GBM algorithm \citep{Goldstein2019}, we conservatively use the original localization uncertainties at this stage. 

We compare the localization of every GRB with all previously observed GRBs. If the circular uncertainty regions with radius $\sigma_{\text{tot}}$ overlap, we place the pair in the L sample. If the uncertainty regions are separated by more than $10^{\circ}$, we place them in the NL sample. The extra separation gives us further confidence that the NL sample contains no lenses. 
We note that GBM uncertainty regions are typically not circular, often being better described by ellipses. Given the conservative nature of these cuts, we expect the use of circular uncertainty regions to have a negligible impact on the final results. We consider more accurate, non-circular uncertainty regions in Section~\ref{sec:timeResolvedSpectra}, where we also generate new response files for GRBs that have better localizations from other instruments.

\textbf{Duration:}
The lensing is not expected to significantly change the duration of a GRB. However, the observed duration may be affected by several uncertainties, including a low signal-to-noise ratio (SNR), background properties, which may also further impact the SNR, as well as the satellite position. In the latter case the observation may start late or stop early due to occultation by the Earth or entering the south-Atlantic-anomaly. To account for these effects, we impose the conservative requirement that $T_{90}$ should differ by less than a factor of $5$ in lens candidate pairs. 

\textbf{Spectra:}
The last hard cut is based on spectra. We use two common empirical functions to assess the spectra; the Band function and a cutoff power law (the Comptonized model in the {\it Fermi}~GBM spectral catalog, \citealt{2016ApJS..223...28N}). Although these functions may not capture all spectral features, similar spectra will yield similar fits. Further, we only consider the model parameters pertaining to the low-energy slope and peak energy of the spectrum, since these tend be more well constrained than the high-energy slopes.
GRB pairs where the $2\sigma$ confidence intervals of both parameters overlap in the time-integrated or peak-flux spectra for at least one of the models are kept in the L sample.  
For this part of the analysis we use the spectral parameters available through the online \textit{Fermi} spectral catalog\footnote{Available at \url{https://heasarc.gsfc.nasa.gov/FTP/fermi/data/gbm/bursts/}.} \citep{Gruber:2014cx,vonKienlin:2014dt,2016ApJS..223...28N}. Only bursts with entries in the spectral catalog are eliminated this way (these entries were missing for most bursts from the second half of 2018 and onward at the time of our analysis). This is a conservative cut that helps reduce the number of pairs that we need to compare further. 

\textbf{Final sample:}
The three cuts are performed in order of increasing computational requirement, \ie ~$T_{90}$, spectral parameters, and position. Naturally, the order of the cuts does not matter for the final L sample. In this order, the three cuts remove $\sim 1.8\cdot 10^{6}$, $\sim 6\cdot 10^{5}$, and $\sim 1.2\cdot 10^{6}$ pairs from the L sample, respectively. This leaves about $1.2 \cdot 10^{5}$ GRB pairs in the L sample to investigate further. We summarize the cuts in Table~\ref{tab:cuts}. 
For comparison, the NL sample contains about $1.1 \cdot 10^{6}$ pairs. This sample has a different cut on position, as described above, but the same cuts on duration and spectra as the L sample. 

The cuts described above are all conservative. We have tested multiple variations of the cuts, with different limits on the localization, spectral parameters and relative $T_{90}$, and find that we obtain largely the same candidates after the light curve comparison in Section~\ref{sec:lightcurves}. 
We stress that there are no cuts based on the relative flux of the GRB pairs. In simple spherically symmetric lens models, the later GRB is expected to be fainter (e.g., \citealt{Mao1992}), but this does not hold for more realistic scenarios. Finally, we note that there are 96 pairs in the initial sample with redshifts that agree to within $5\%$ of each other, but that none of these pairs survive the hard cuts.

\subsection{Cross correlation of light curves} \label{sec:lightcurves}
The next step of the analysis is to assess the similarity of light curves for the GRB pairs that remain after the initial cuts. We produce light curves from the TTE data using two different bin sizes, $0.5$ and $0.05$~s, and perform the analysis for both. The smaller bin size is particularly useful for short GRBs, where a larger bin size will yield featureless light curves. We construct light curves over the longest possible time interval for each GRB, something which is mainly limited by the available detector response, and also subtract the background. The background was determined by fitting first-order polynomials in \textit{3ML}. We sum the background-subtracted light curves from the brightest NaI detectors for each burst (usually three detectors), as listed in the \textit{Fermi}~GBM spectral catalog. Using multiple detectors helps compensate for different observing conditions for single detectors. We do not include the BGO detectors since they make very little difference for the time series analysis and because information on which BGO detector to use is not available for the most recent GRBs.

In order to assess the similarity between two light curves, we consider the cross correlation (CC). We use the implementation of the CC in one dimension for discrete, real-valued functions $a[i]$ and $b[i]$, normalized such that the maximum value of the CC is unity;
\begin{align*}
    CC[n] = \sum _{i=0}^{N-1}\frac{{a[i]}b[i+n]}{N \cdot \sigma_\text{a}\sigma_\text{b}}.
\end{align*}
Here, $a$ and $b$ represent the binned light curves, $n$ is the relative displacement of $a$ and $b$, and $\sigma_\text{a}$ and $\sigma_\text{b}$ are the standard deviations of $a$ and $b$, respectively. We have also assumed that both $a$ and $b$ are of length $N$. This is important because the relative lengths of the light curves could otherwise impact the maximum value of the resulting CC. 
For each burst pair we therefore trim the longer light curve such that it is of the same length as the shorter light curve, while still containing the relevant emission episode as defined by the $T_{90}$.

For each pair we use the maximum of the CC (henceforth denoted CC$_{\text{max}}$) as the measure of the similarity between the light curves. It is obvious that CC$_{\text{max}}$ will tend to increase with bin size and that one should not compare values of CC$_{\text{max}}$ constructed from light curves with different binning. 
We also find that using background-subtracted light curves is important for getting reliable values of CC$_{\text{max}}$.  When calculating the CC for non-background subtracted light curves, we find a large number of pairs that have artificially high CC$_{\text{max}}$ due to similarly varying backgrounds. We therefore base all our analysis and results on background-subtracted light curves.

In order to assess the results we also performed a simulation study. The simulations are described in appendix~\ref{appendix:simulations}. In Figure~\ref{fig:CCdistr_L_SL} we show the CC$_{\text{max}}$ distributions of the L sample for both time binnings together with the corresponding simulated distributions. 
The simulated lensing scenario is simplified and does not consider any specific lens model, major variations in observing conditions or biases in the L sample (see appendix~\ref{appendix:simulations} and Section~\ref{sec:Discussion} for further discussion). However, it does provide an estimate of the range of CC$_{\text{max}}$ values that are compatible with lensing. 

\begin{figure}[b] 
\centering
    \includegraphics[width=0.42\textwidth]{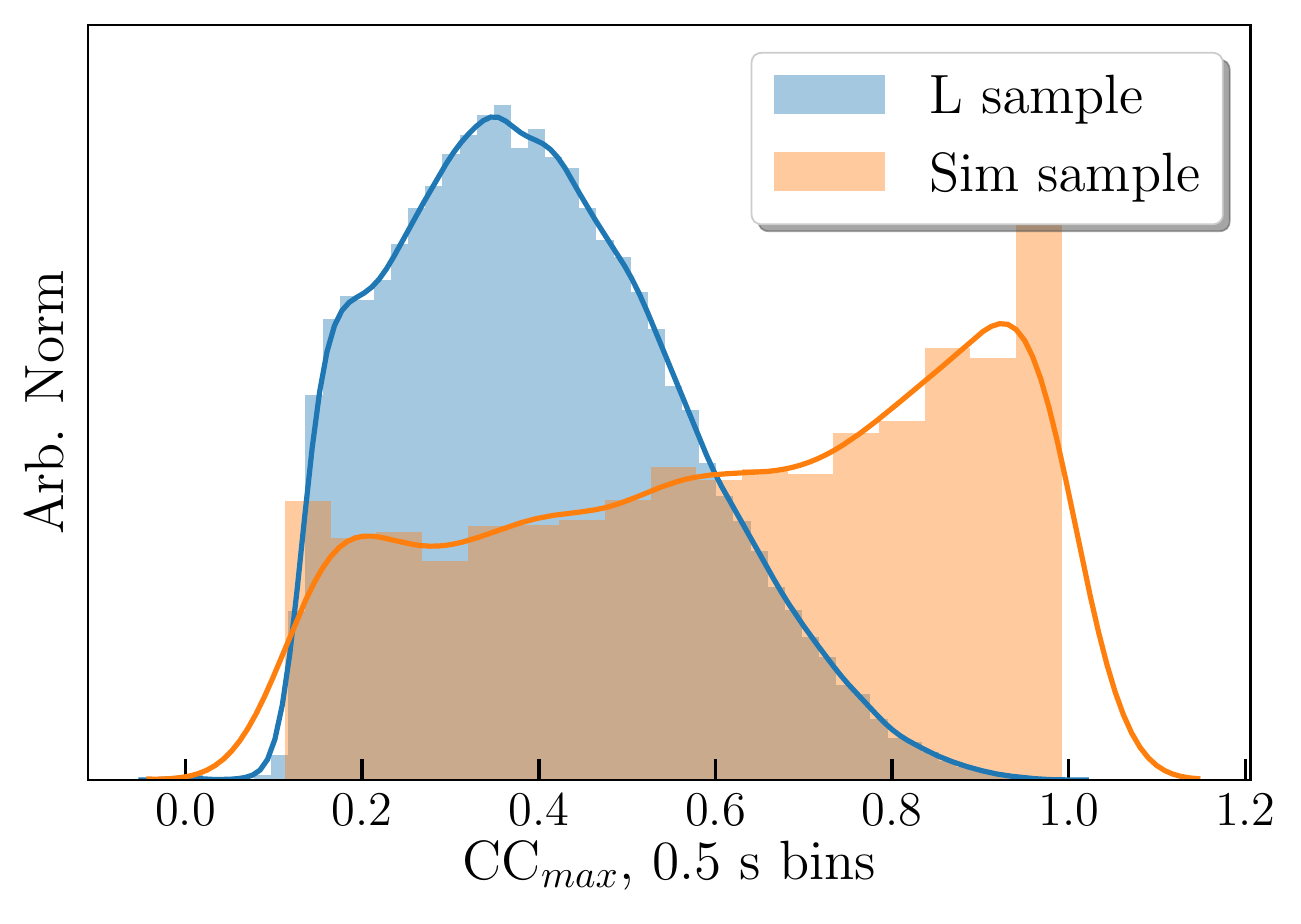}
    \includegraphics[width=0.42\textwidth]{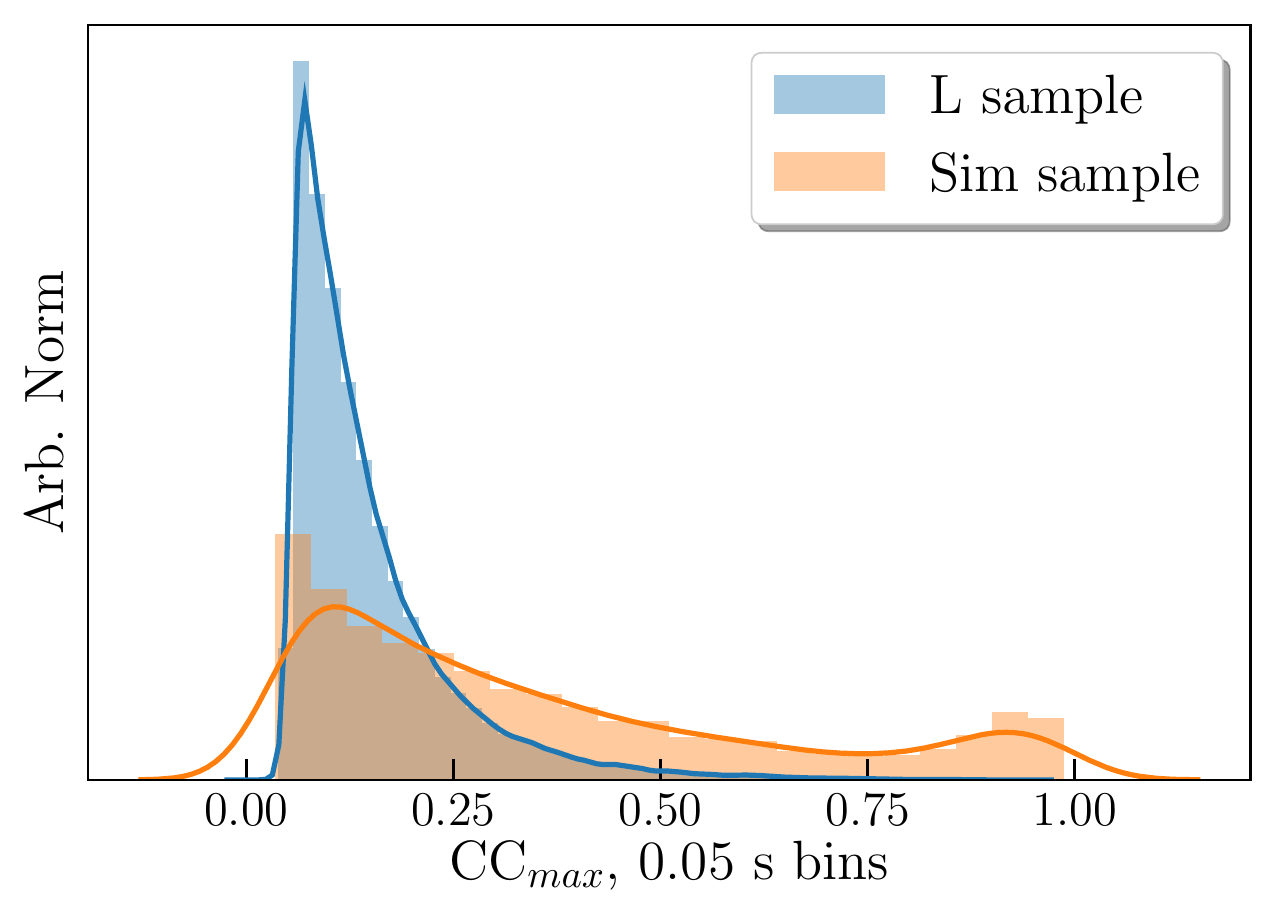}
\caption{Distribution of CC$_{\text{max}}$ for the L sample (blue) and the simulations (orange). The results for time bins of $0.5$ and $0.05$~s are shown in the top and bottom panels, respectively. The histograms for the simulations are constructed from 5500 light curve pairs (see  appendix~\ref{appendix:simulations} for further details about the simulations).} \label{fig:CCdistr_L_SL}
\end{figure}

The distribution of CC$_{\text{max}}$ from the simulations suggests that higher values of  CC$_{\text{max}}$ are increasingly indicative of lensing, as expected. However, even if we assume that the simulations include the most important variations present in lensed GRB light curves, there is no clear value of CC$_{\text{max}}$ to use as a cutoff. 
In addition, while the simulations show that low values of CC$_{\text{max}}$ can be compatible with lensing, the lack of redshift measurements means that very similar light curves are still needed in order to identify convincing lens candidates.
For these reasons, we simply select the burst pairs with the 250 highest values of CC$_{\text{max}}$ for both time bins for further analysis. Due to overlap, this results in 315 unique burst pairs from the L sample that we analyze below.  

\subsection{Time-resolved spectral analysis} \label{sec:timeResolvedSpectra}
The final step is a time-resolved spectral analysis. However, we first refine the sample further. A subset of \textit{Fermi}~GBM bursts have improved localization information available in the online catalog, consisting of sky map probability distributions that account for both the statistical and systematic uncertainties \citep{Connaughton:2015gp}.
These non-circular regions provide more accurate estimates of the localization uncertainties than the $\sigma_{\rm tot}$ used in Section~\ref{sec:samples}.

We inspect the updated localizations, plots of the model spectra, as well as the light curves of all 315 burst pairs, and manually select the 22 most promising candidates for time-resolved spectral analysis. By performing a time-resolved spectral analysis, we obtain significantly better estimates of spectral parameters than what is available through the GBM catalog. 
There are four GRBs in the sample selected for time-resolved spectral analysis that have reported positions from {\it Swift} (GRB~120119A, GRB~161004B, GRB~180703A and GRB~190720A). For these GRBs, we generate new response files for the improved positions using the GBM response generator.\footnote{\url{https://fermi.gsfc.nasa.gov/ssc/data/analysis/gbm/DOCUMENTATION.html}} We find that the new response files have a small impact on the spectral analysis, with the best-fit parameters being consistent within 1$\sigma$.

We perform a Bayesian analysis of the time-resolved spectra with a cutoff power-law using \textit{3ML}. We use a flat prior for the spectral index, $\alpha \sim \mathcal{U}(-10,10)$, and priors that are flat in logarithmic space for the cutoff energy and normalization, $\log x_\mathrm{c} \sim \mathcal{U}(1,1000)$~keV and $\log K \sim \mathcal{U}(10^{-30},10^{3})~\mathrm{keV}^{-1} \mathrm{s}^{-1} \mathrm{cm}^{-2}$, respectively. Further, we sample the posterior using the built-in python implementation \citep{2014A&A...564A.125B} of MultiNest \citep{10.1111/j.1365-2966.2009.14548.x}, using 500 samples. For a few spectra we use different limits on the priors and a different number of samples in order to achieve convergence.

Since the trigger times may vary somewhat relative to the overall light curve shapes, we align the bursts by shifting the starting time of the second GRB. The alignment was done by eye, but it essentially retrieves the optimal time lag found by the CC analysis. The shift allow us to use the same time bins for the spectral analysis of both GRBs. We define these time bins by running the Bayesian blocks algorithm \citep{2013ApJ...764..167S} on the first GRB. 
Background spectra for all time intervals were created as described in Section~\ref{sec:lightcurves}.

We finally compare the resulting posteriors in each time bin. In the ideal lens scenario we expect these to overlap to a large degree in all bins that have a significant signal. The time-resolved analysis allows us to eliminate all the remaining candidates. In the next Section we present the most interesting cases from this analysis.

%%%%%%%%%%%%%%%%%%%%%%%%%%%%%%%%%%%%%%%%%%%%%%%%%%%%%%%%%%%%%%%%%%%%%%%%%%%

\section{Analysis of final candidates} \label{sec:FinalLensCands}  
\begin{figure*}
    \centering
    \includegraphics[width=0.30\textwidth]{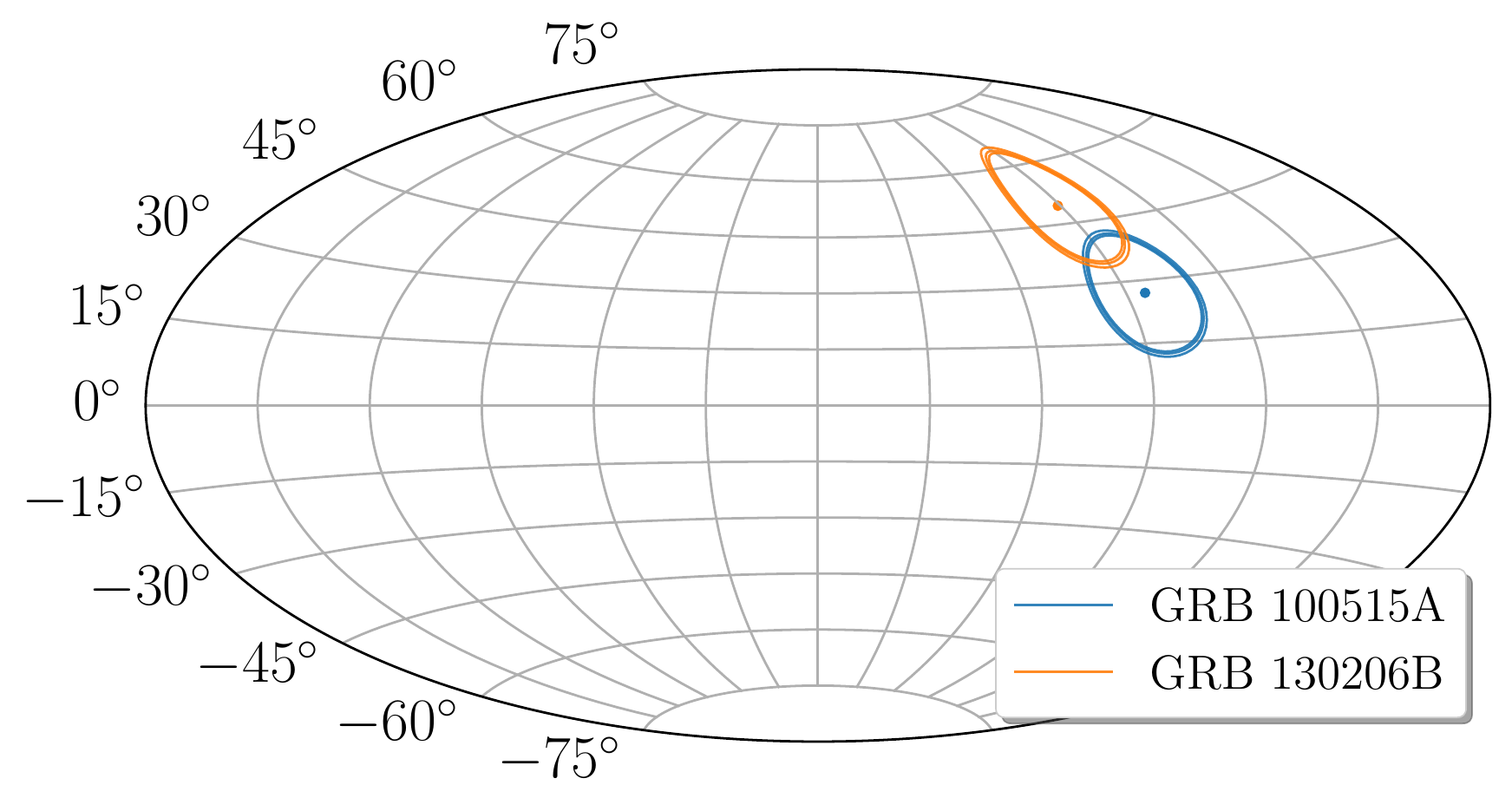}
    \includegraphics[width=0.30\textwidth]{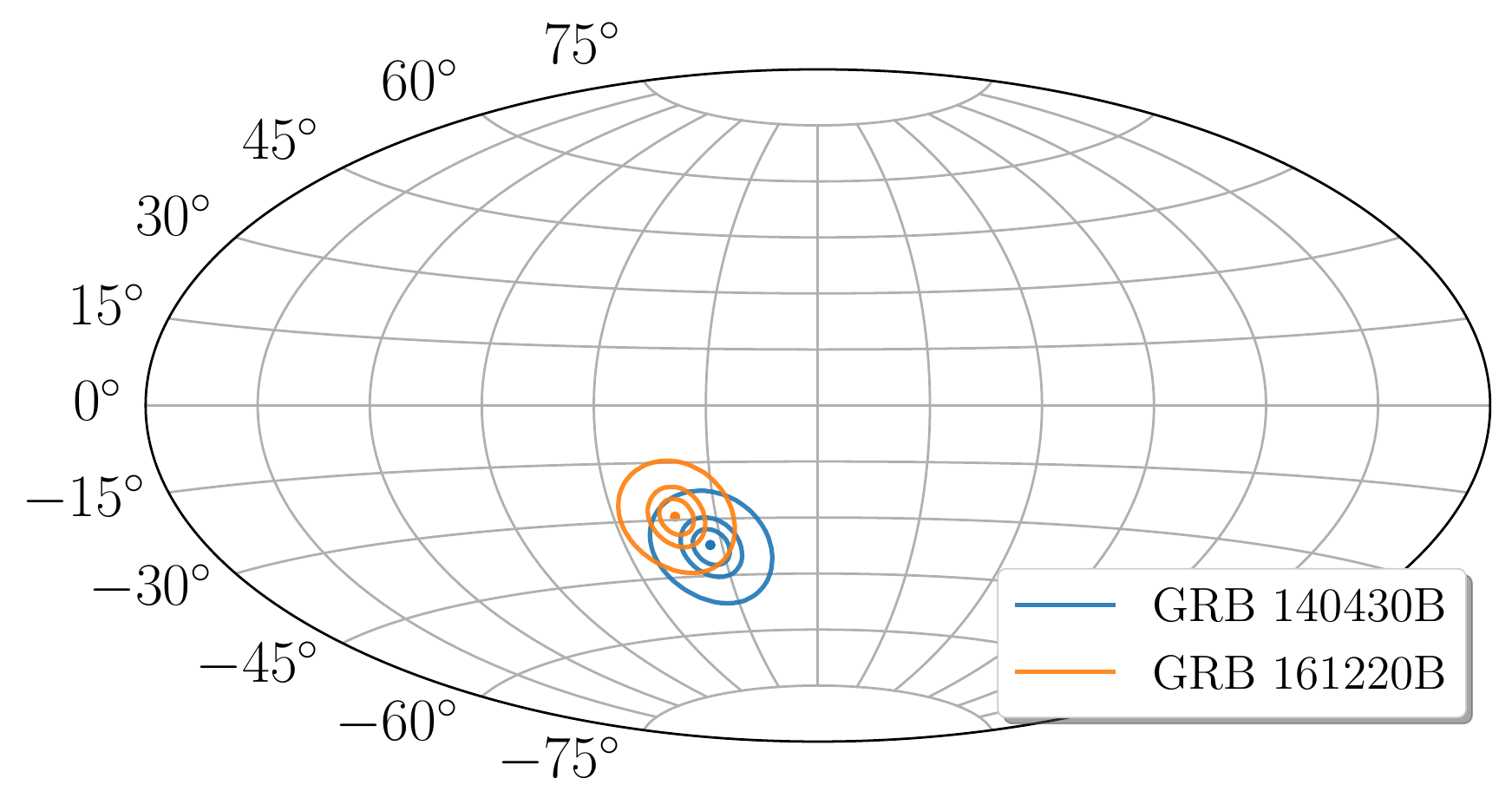}
    \includegraphics[width=0.30\textwidth]{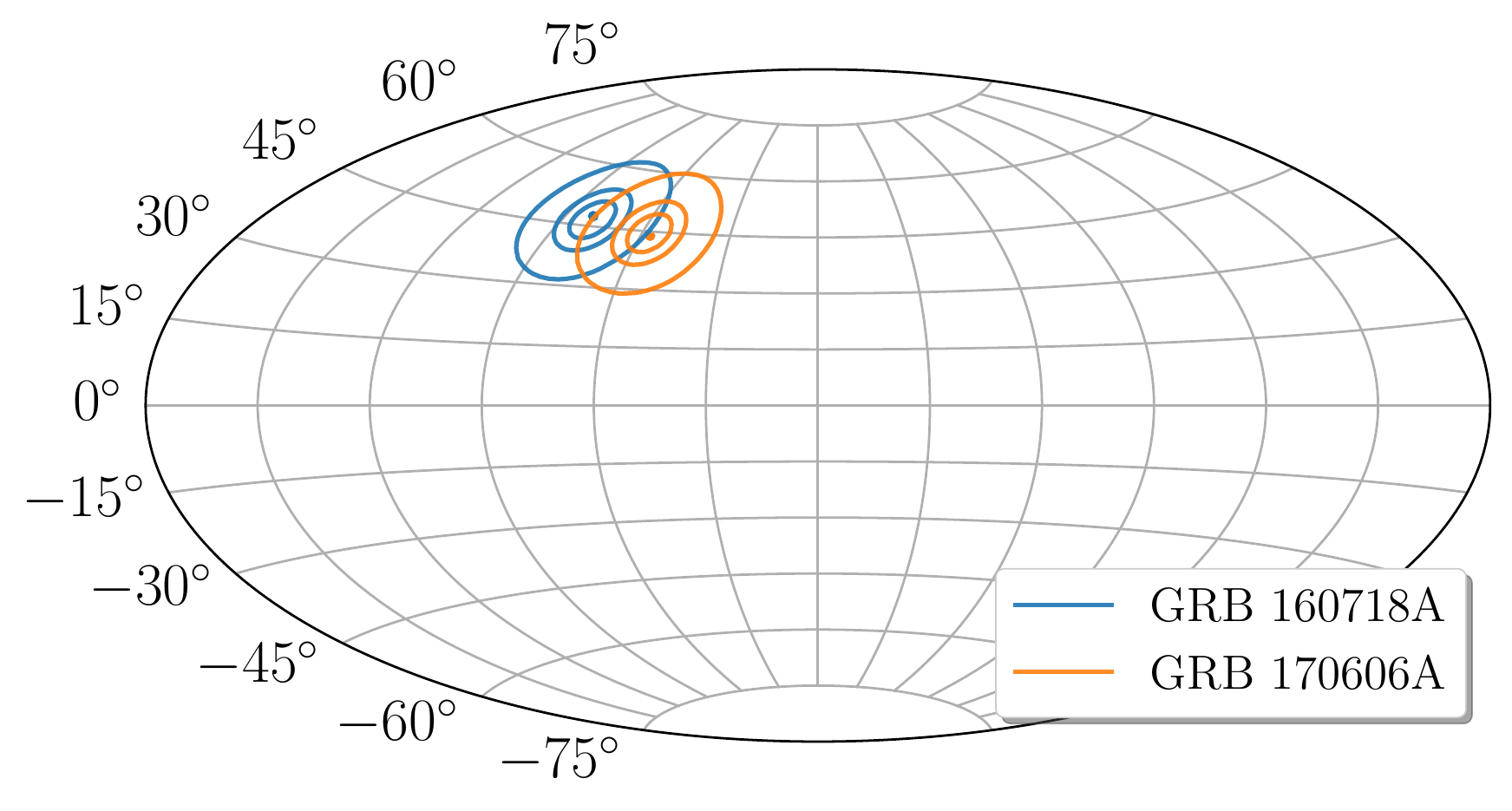}
    \caption{Sky localizations for GRB~100515A-GRB~130206B, GRB~140430B-GRB~161220B, and GRB~160718A-GRB~170606A. The lines indicate the 1, 2, and 3$\sigma$ uncertainty regions, respectively. For GRB~100515A and GRB~130206B, the uncertainty regions are calculated as described in Section~\ref{sec:samples}. Note that the statistical errors are  small for these two GRBs, which means that the three confidence levels of each burst overlap almost completely and that the uncertainty is dominated by the systematic uncertainty. For the other four GRBs, \textit{Fermi}~GBM supplies uncertainty regions that include both the statistical and systematic uncertainties. These regions happen to be approximately circular for these GRBs.}
    \label{fig:positions}
\end{figure*}

\begin{figure*}
\centering
    \includegraphics[width=0.40\textwidth]{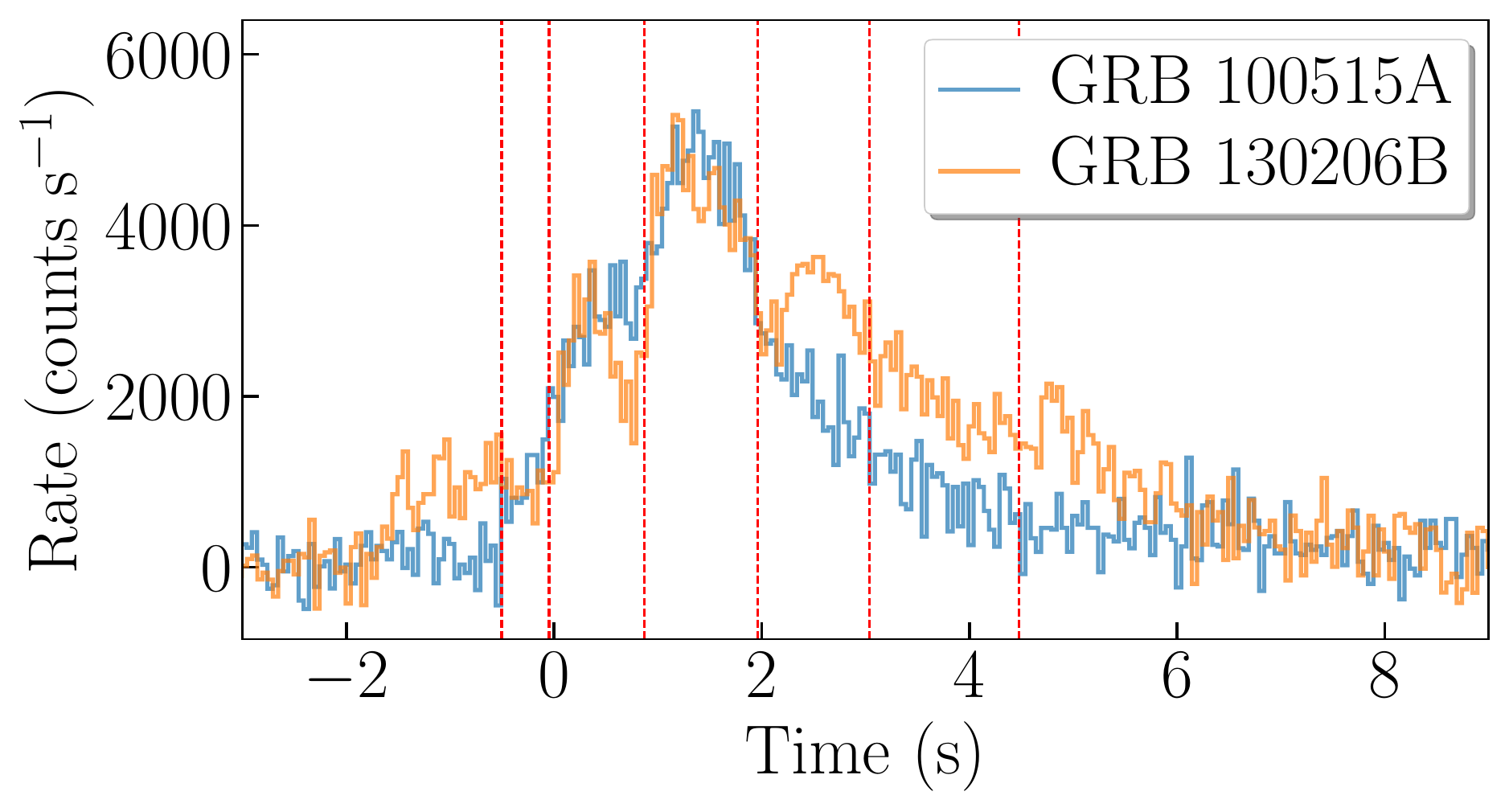}
    \includegraphics[width=0.40\textwidth]{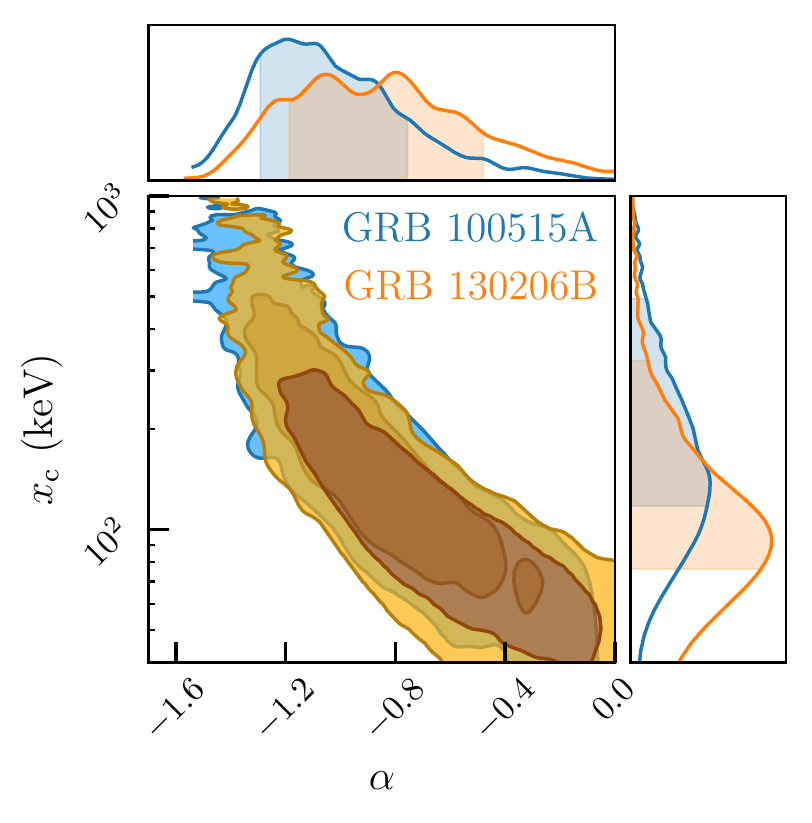}
    \includegraphics[width=0.40\textwidth]{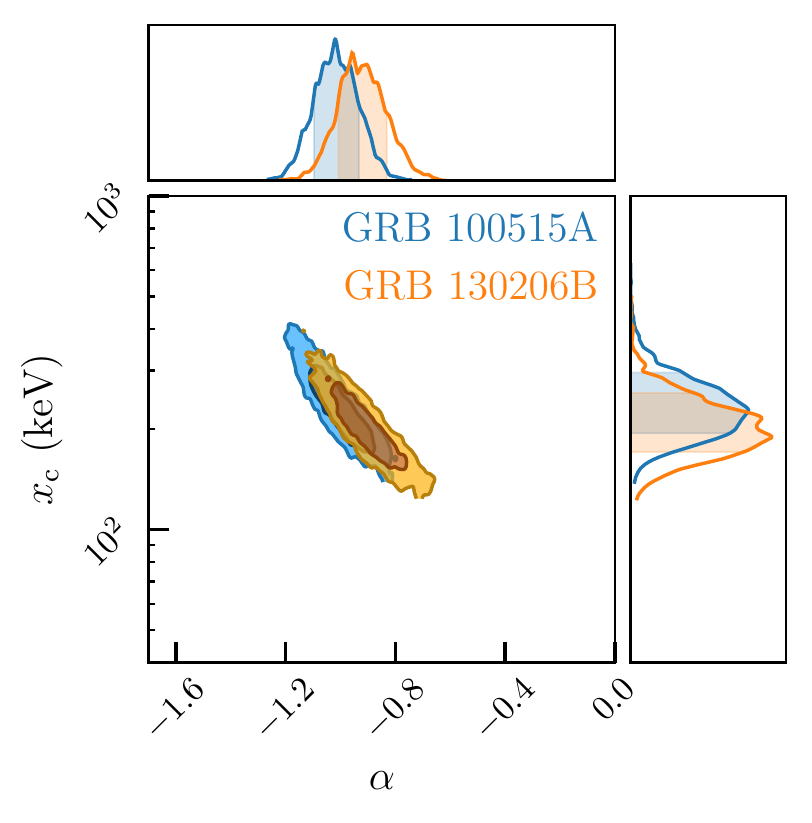}
    \includegraphics[width=0.40\textwidth]{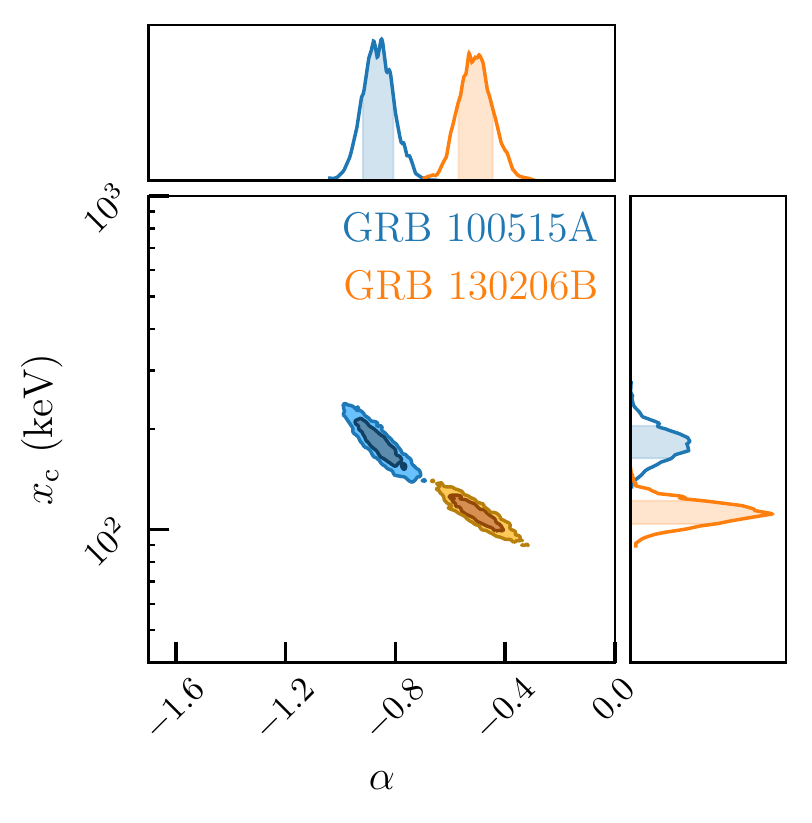}
    \includegraphics[width=0.40\textwidth]{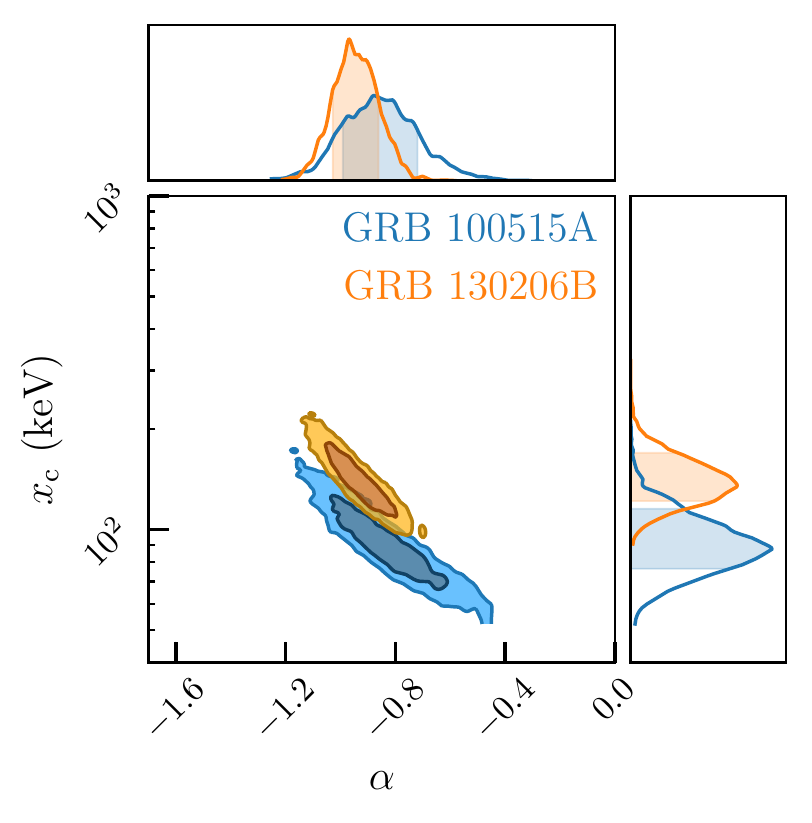}
    \includegraphics[width=0.40\textwidth]{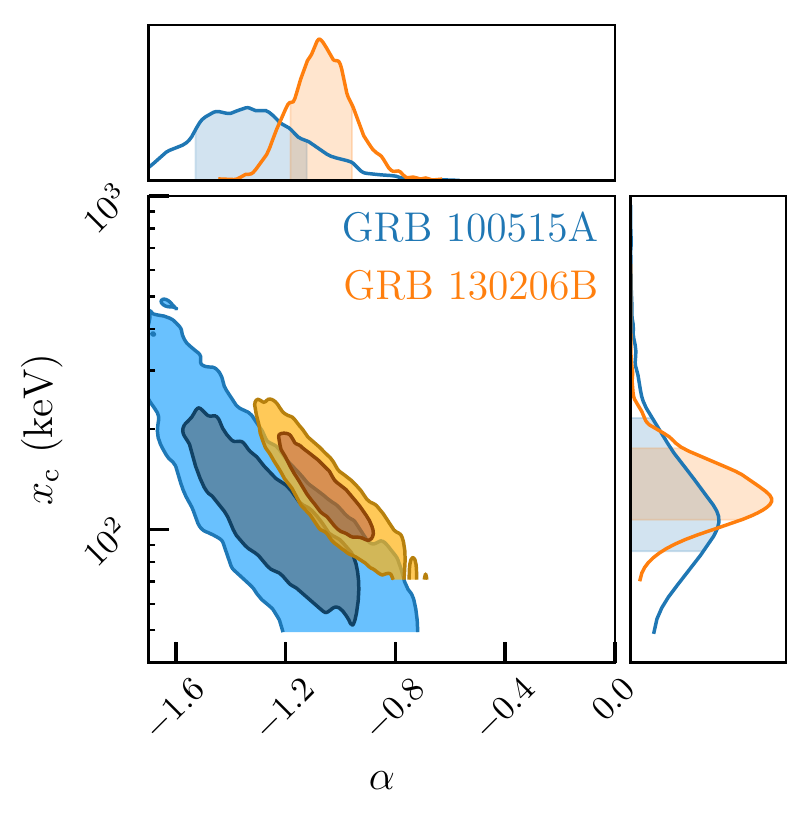}
    \caption{Light curves of GRB~100515A and GRB~130206B together with the posterior distributions from fitting the time-resolved spectra with a cutoff power law. The red lines in the light curves indicate the edges of the time bins used for the spectral analysis. The dark and light shaded regions represent the 68 and 95 \% credible regions, respectively. }
    \label{fig:100515-130206}
\end{figure*}

\begin{figure*}
\centering
    \includegraphics[width=0.40\textwidth]{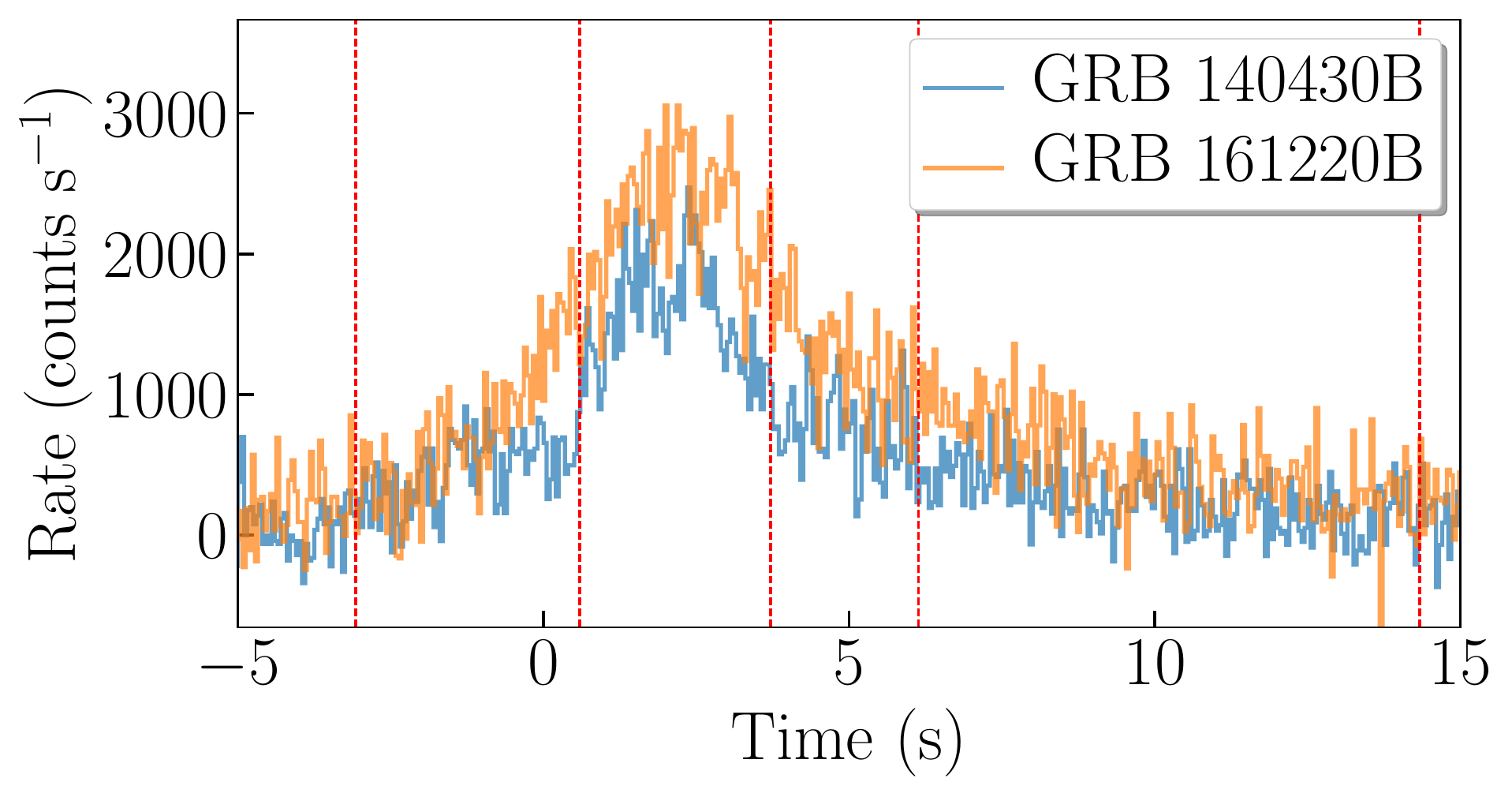}
    \includegraphics[width=0.40\textwidth]{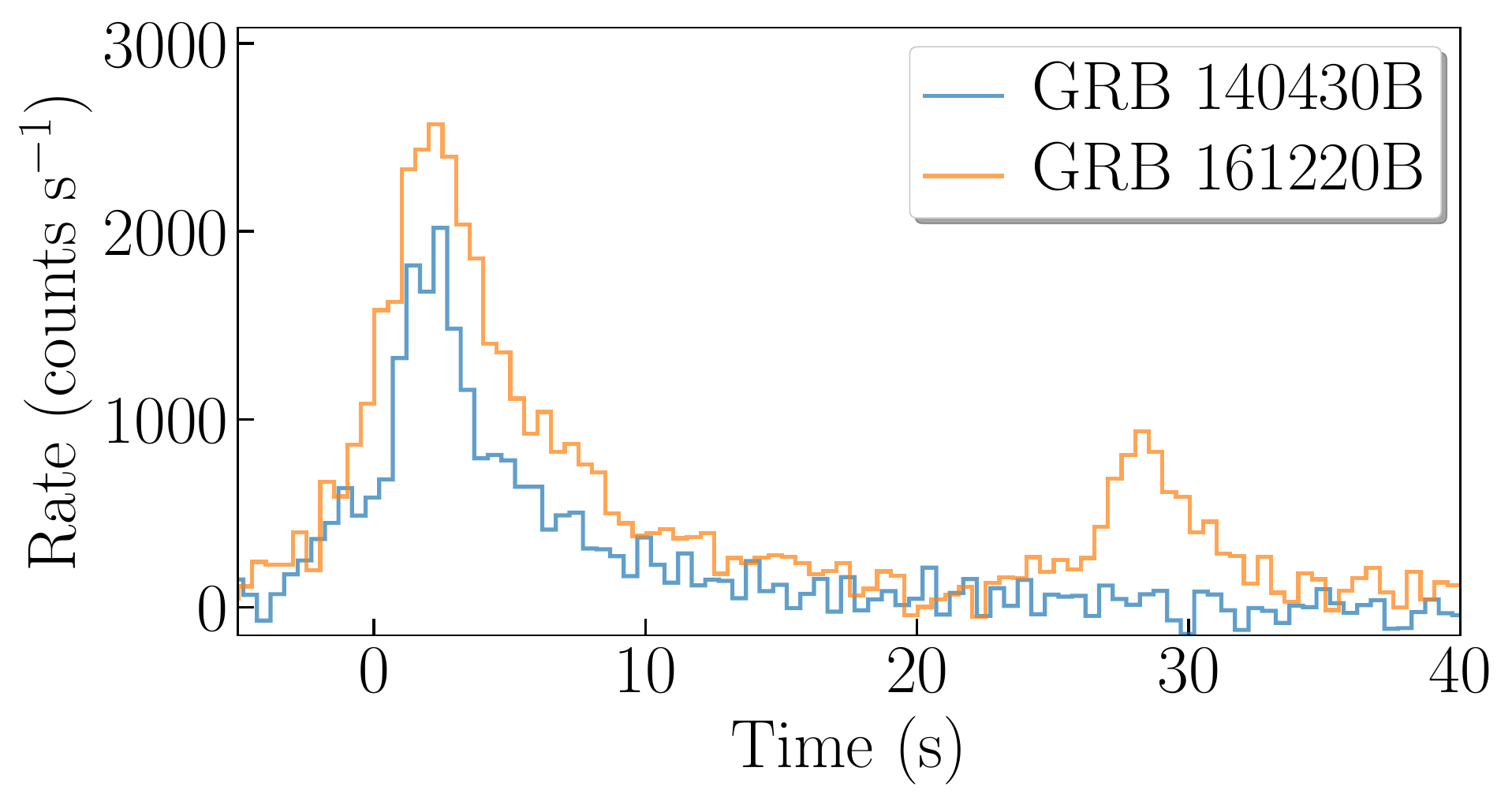}
    \includegraphics[width=0.40\textwidth]{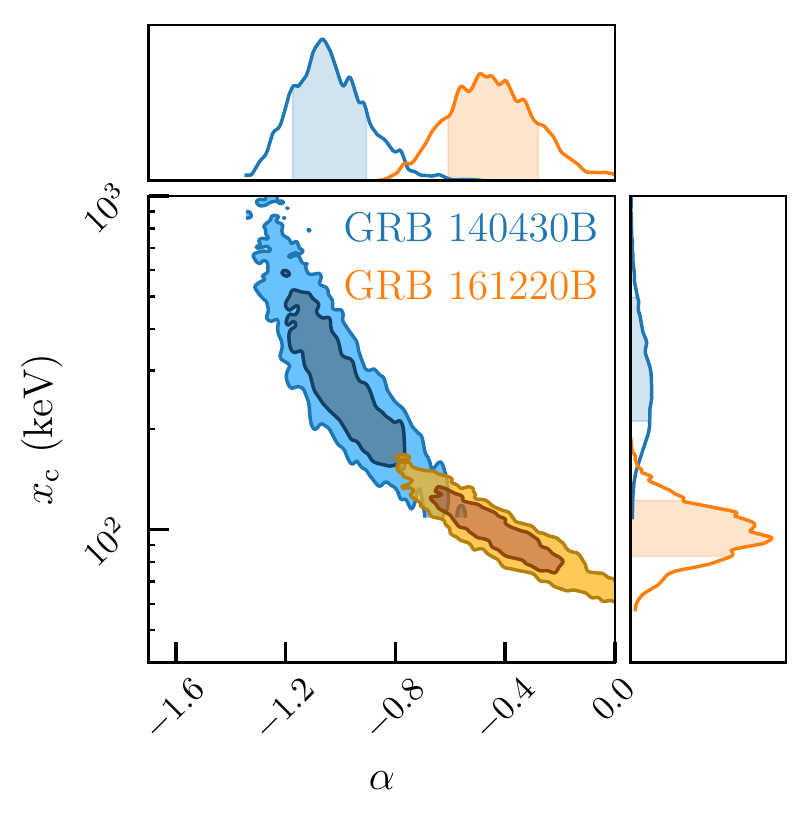}
    \includegraphics[width=0.40\textwidth]{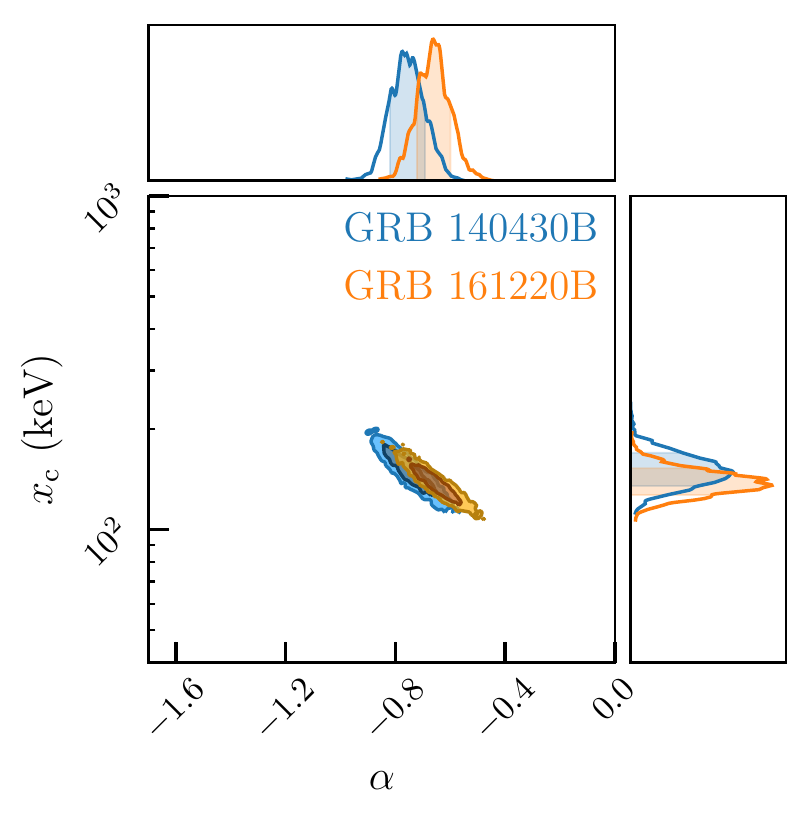}
    \includegraphics[width=0.40\textwidth]{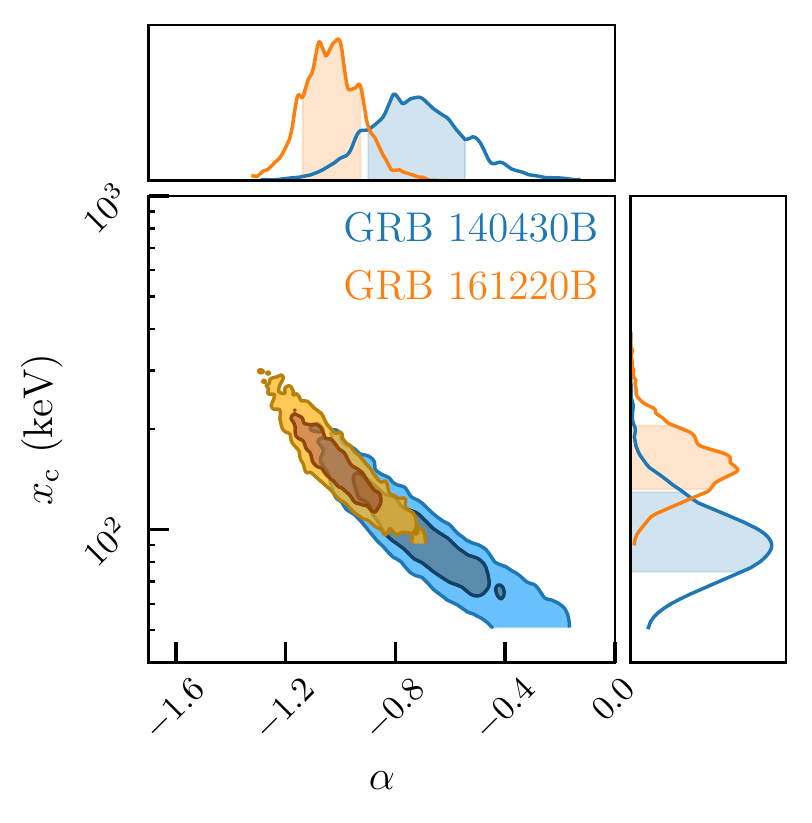}
    \includegraphics[width=0.40\textwidth]{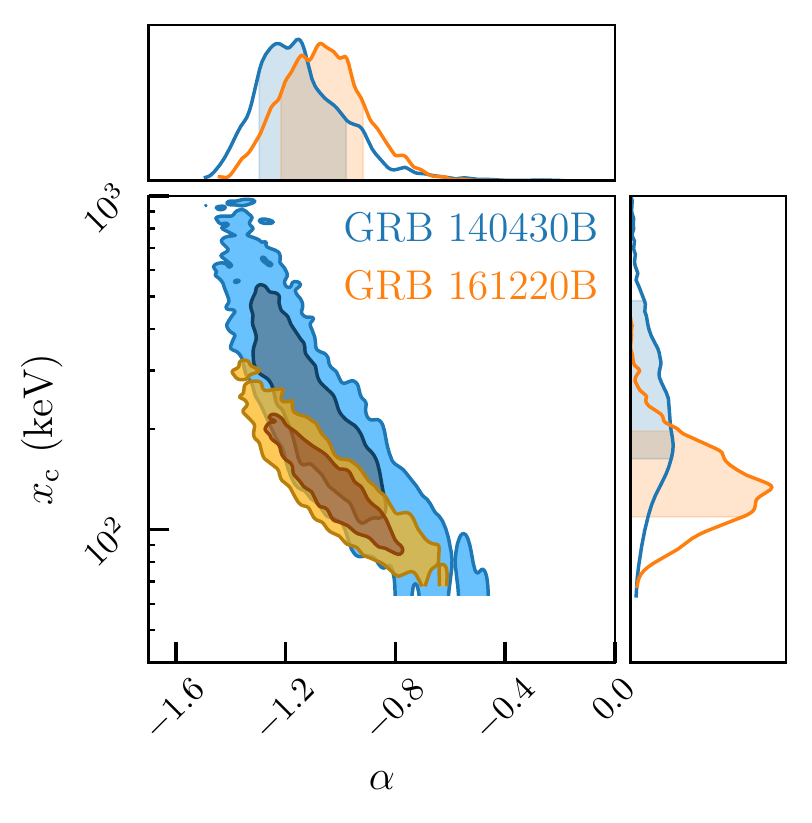}
    \caption{Same as Figure~\ref{fig:100515-130206}, but for GRB~140430B and GRB~161220B. Note that GRB~161220B has an additional peak in the light curve at $\sim 30$~s (top right panel), which casts doubt on these bursts as a lensed pair.}
    \label{fig:140430-161206}
\end{figure*}

\begin{figure*}
\centering
    \includegraphics[width=0.40\textwidth]{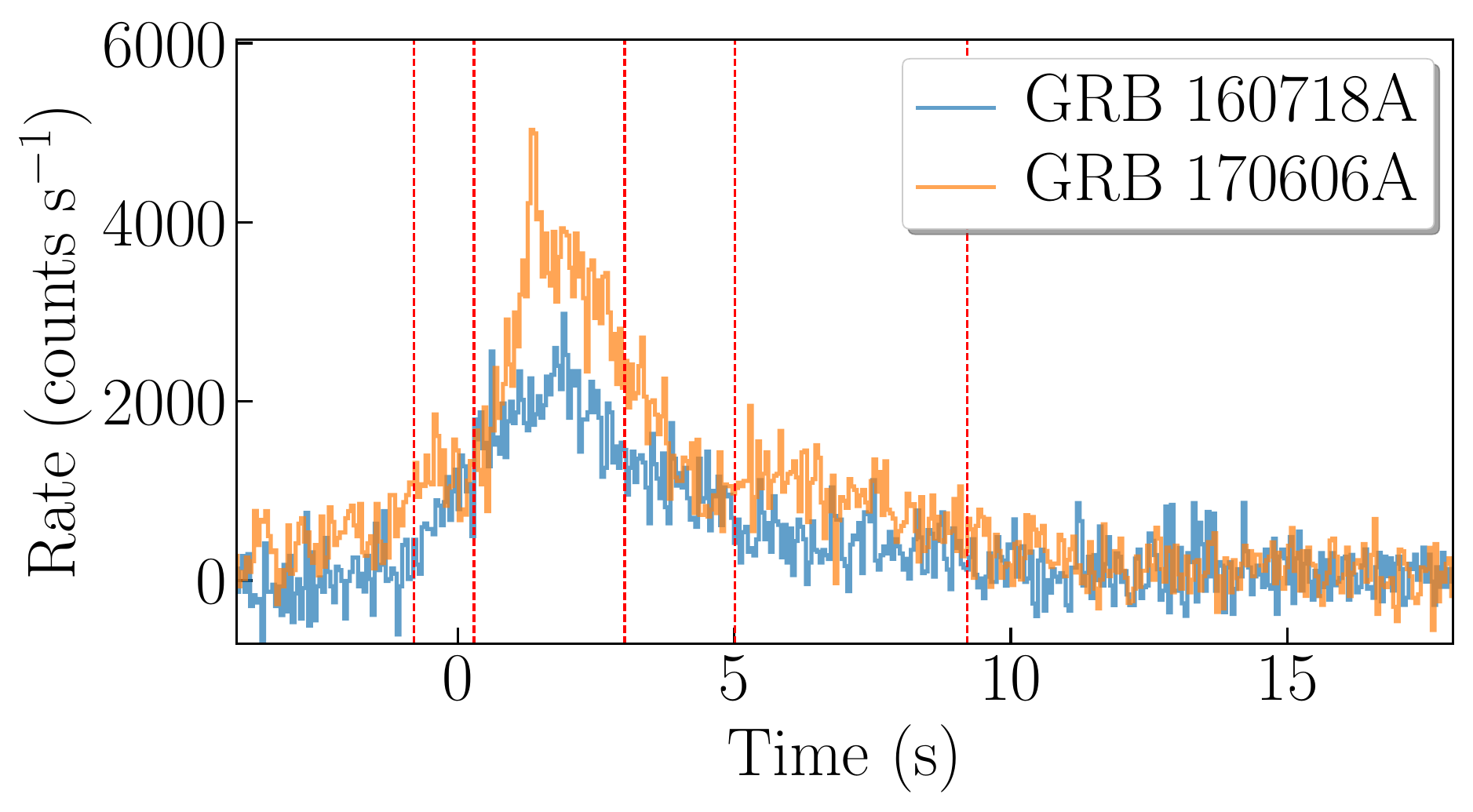}
    \includegraphics[width=0.40\textwidth]{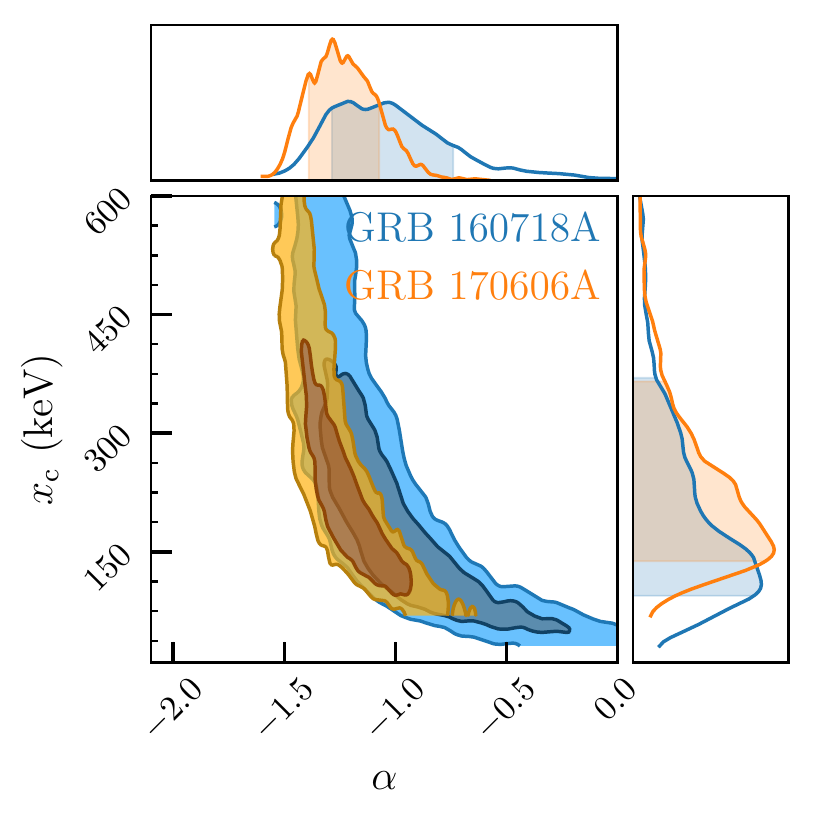}
    \includegraphics[width=0.40\textwidth]{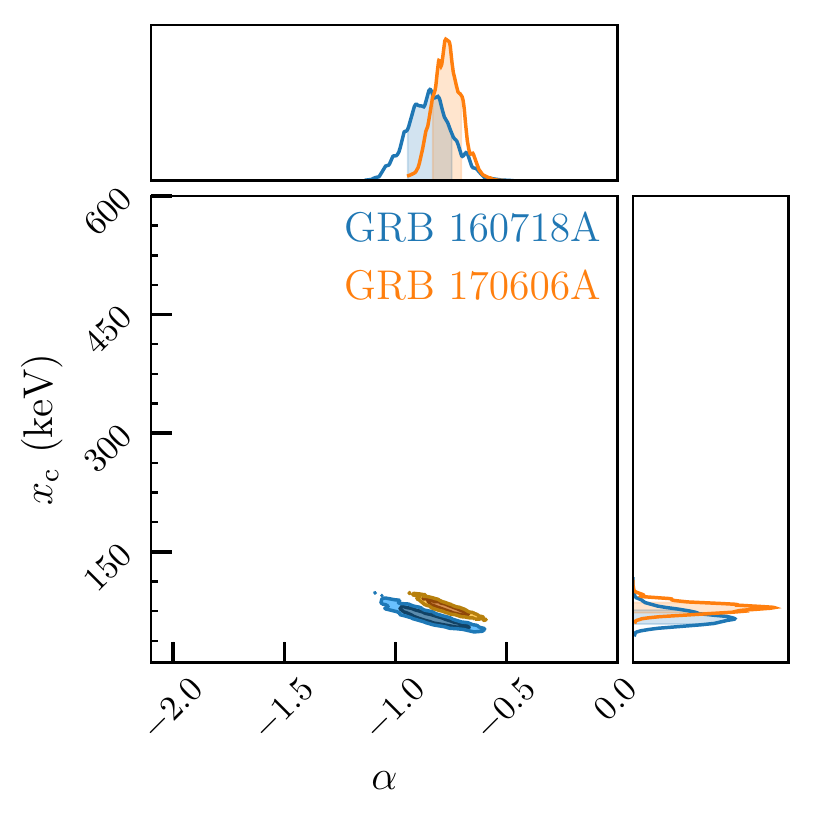}
    \includegraphics[width=0.40\textwidth]{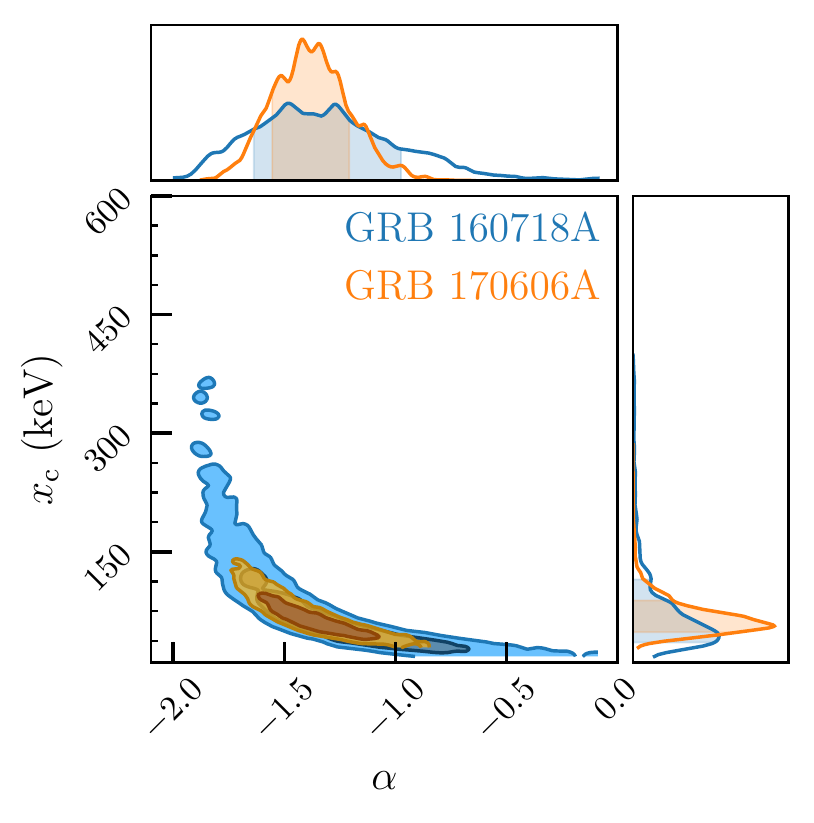}
    \includegraphics[width=0.40\textwidth]{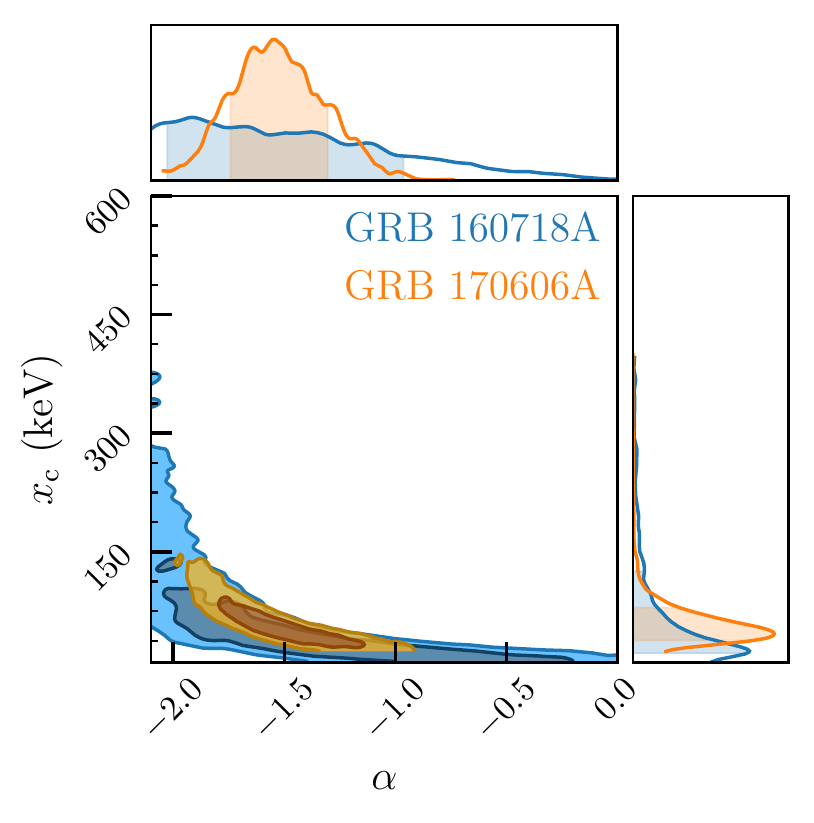}
    \caption{Same as Figure~\ref{fig:100515-130206}, but for GRB~160718A and GRB~170606A. Note the discrepancy between the posteriors in the second time bin.}
    \label{fig:160718-170606}
\end{figure*}

In Table~\ref{tab:cuts} we summarize the number of GRB pairs present in the L sample after each cut, starting from the number of unique pairs constructed from 2712 GRBs. Following the steps described in Sections~\ref{sec:samples}-\ref{sec:timeResolvedSpectra}, we are left with no convincing candidates for gravitationally lensed GRBs. For illustrative purposes we present the three most interesting candidates below. These are GRB~100515A-GRB~130206B, GRB~140430B-GRB~161220B, and GRB~160718A-GRB~170606A. We also use the last pair as a case study to investigate the effects of observational uncertainties. The sky positions for all three pairs are shown in Figure~\ref{fig:positions}, while Figures~\ref{fig:100515-130206}-\ref{fig:160718-170606} show the light curves and evolution of spectral parameters.

\begin{deluxetable}{ll}
\tablenum{1}
\tablecaption{Number of GRBs in the L sample after cuts. The first three are hard cuts, described in Section~\ref{sec:samples}, while the last three are soft cuts, described in Sections~\ref{sec:lightcurves}, \ref{sec:timeResolvedSpectra} and \ref{sec:FinalLensCands}.} \label{tab:cuts}
\tablewidth{0pt}
\tablehead{
Type of cut & GRB pairs remaining
}
% \decimalcolnumbers
\startdata
No cut (initial sample) & $3~676~116$\\
$T_{90}$ & $1~901~542$ \\
Time-integrated spectra  & $1~292~816$\\
Position & $116~335$ \\
\hline
CC$_{\text{max}}$ & 315 \\
Refined localization  &  \\
\& manual selection  & 22 \\
Time-resolved spectra & 0 \\
\enddata
% \tablecomments{Summary of lens candidates}
\end{deluxetable}

GRB~100515A-GRB~130206B has a CC$_{\text{max}}$ of $0.89$ and $0.74$ for the $0.5$ and $0.05$~s bins, respectively. However, Figure~\ref{fig:100515-130206} shows that  the light curves look significantly different by eye, particularly at $2$ -- $5$~s. Furthermore, the spectra differ significantly at the peak of the light curve. This is despite the fact that these bursts passed the time-averaged spectral cuts described in Section~\ref{sec:samples}. Finally, as can be seen in Figure~\ref{fig:positions}, the localization uncertainty regions overlap only marginally, and due to the fact that we have used conservative estimates of the systematic uncertainty of the GBM localization. Neither of these bursts have improved localization available. This pair is not considered to constitute a lensing event.

GRB~140430B-GRB~161220B, seen in Figure~\ref{fig:140430-161206}, show good agreement in the main emission episode, both in terms of light curves and overlapping spectral parameters. However, this pair is rejected on the basis of GRB~161220B having an additional peak in the light curve $\sim 20$~s after the main episode, which is not present in GRB~140430B (Figure~\ref{fig:140430-161206}, second panel). There is a possibility that the second peak in GRB~161220B is the result of millilensing, where lensing by, 
e.g., massive black holes ($M\gtrsim 10^6~{\rm M_{\odot}}$) leads to repeating emission episodes within the GRB \citep{Nemiroff2001}. However, fitting the time integrated spectra of the two light curve peaks, we find that the posteriors have almost no overlap at the $95$~\% level, with the second peak having a softer spectrum. Furthermore, this burst pair exhibits simple light curve shapes with few time bins to analyze, making it a less compelling case than an overlap between more complex light curves.

The pair GRB~160718A-GRB~170606A has CC$_{\text{max}}$ of $0.91$ and $0.63$, and shows good overlap of the spectra in three out of four bins. However, the second time bin in Figure~\ref{fig:160718-170606} shows a small, but significant, discrepancy between the posteriors, as well as small differences between the light curves. 
Below, we investigate whether these differences may be caused by observational uncertainties and/or the flux difference expected from a lensing scenario. We consider the CC$_{\text{max}}$ and visual appearance of the light curves in Section~\ref{sec:crossCorrelation}, and the spectral properties in Section~\ref{sec:spectralAnalysis}. 
%%%%%%%%%%%%
We find that the differences between the light curves and the observed values of CC$_{\text{max}}$ are in fact consistent with lensing, but that the spectral differences are not.
However, even if there were better spectral overlap in the second time bin, the light curves of GRB~160718A-GRB~170606A are rather featureless and the number of analyzed time bins few, which makes this a weakly compelling case at best. Finally, although the localization cannot be used to confidently rule out a lensing scenario, there is essentially no overlap of the 2$\sigma$ contours (Figure~\ref{fig:positions}, right panel), further reducing the likelihood of lensing. We thus conclude that GRB~160718A-GRB~170606A is unlikely to be an example of gravitationally lensed GRBs.

\subsection{Impact of observational uncertainties on light curves and cross correlations} \label{sec:crossCorrelation}
As discussed in Section~\ref{sec:lightcurves}, the value of CC$_{\text{max}}$ is in itself not an adequate measure of the probability of lensing.
However, for a given burst pair we can use simulations to quantify how likely the observed CC value is under a specific lensing scenario. This technique can be used to significantly improve the power of CC$_{\text{max}}$ as a tool to identify or reject lensed GRB pairs. To illustrate this, we consider the case of GRB~160718A-GRB~170606A, which is the most promising candidate described above. 

\begin{figure}
    \centering
    \includegraphics[width=0.45\textwidth]{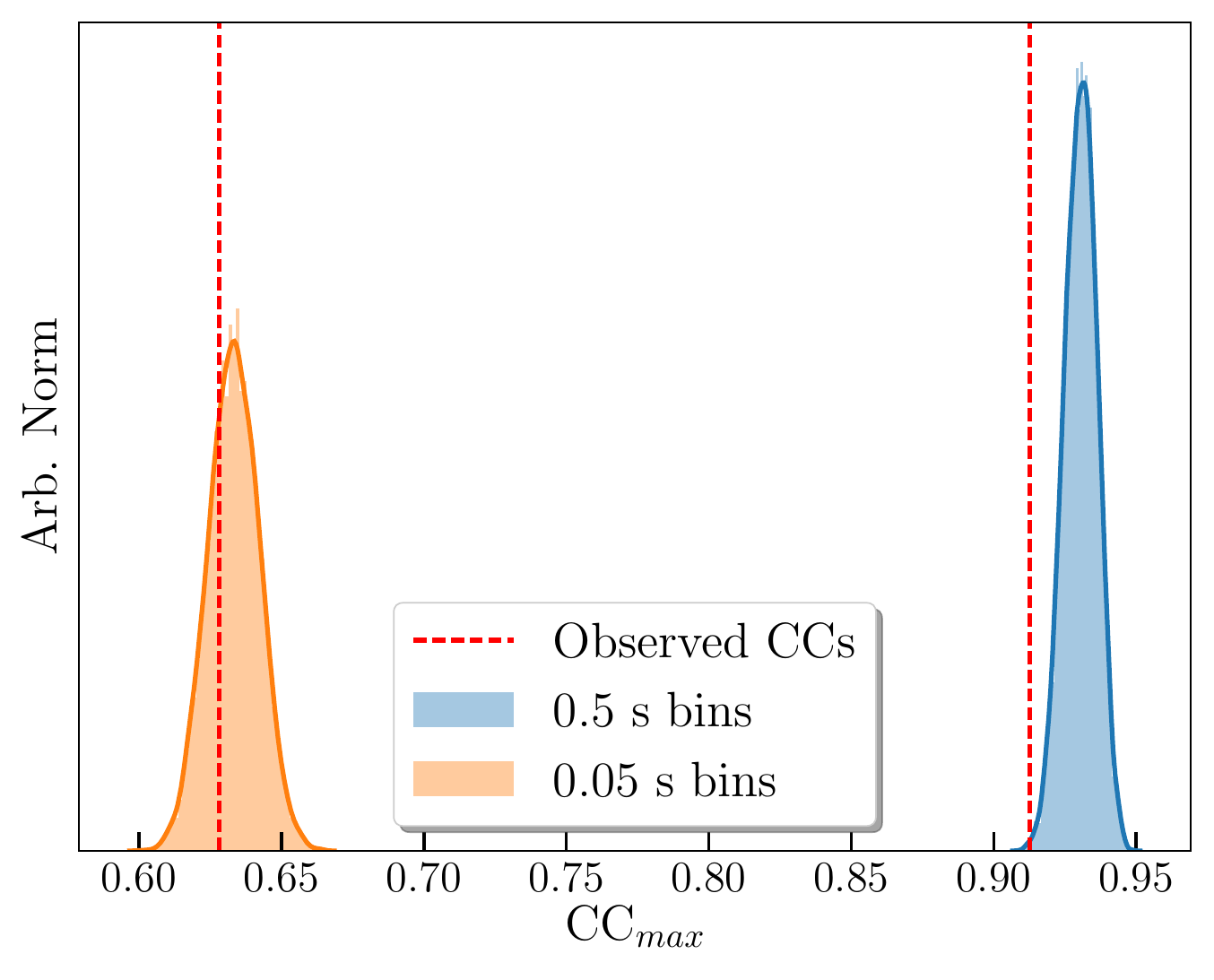}
    \caption{Distributions of CC$_{\text{max}}$ values expected from an ideal lensing scenario of GRB~170606A, assuming a flux change to the level of GRB~160618A. Results for the $0.5$ and $0.05$~s bins are shown in blue and orange, respectively. The dashed red lines show the observed values of CC$_{\text{max}}$ between GRB~170606A and GRB~160618A.}
    \label{fig:CCsim170606}
\end{figure}

We simulate light curves from GRB~170606A (on the basis of it being the brighter one), using the methods described in appendix~\ref{appendix:simulations}, but with the background and signal flux levels set to those of GRB~160718A. The value of CC$_{\text{max}}$ is then calculated between each simulated light curve and the original light curve. In Figure~\ref{fig:CCsim170606}, we present the resulting CC$_{\text{max}}$ distributions for the two time bins together with the observed values of CC$_{\text{max}}$. The observed CC$_{\text{max}}$ from the $0.05$~s bins is fully consistent with the simulated distribution, while there is some tension in the case of the $0.5$~s time bins. Considering the idealized nature of the simulations, these results suggest that we cannot reject the lensing hypothesis for this burst pair.

\begin{figure}
    \centering
    \includegraphics[width=0.45\textwidth]{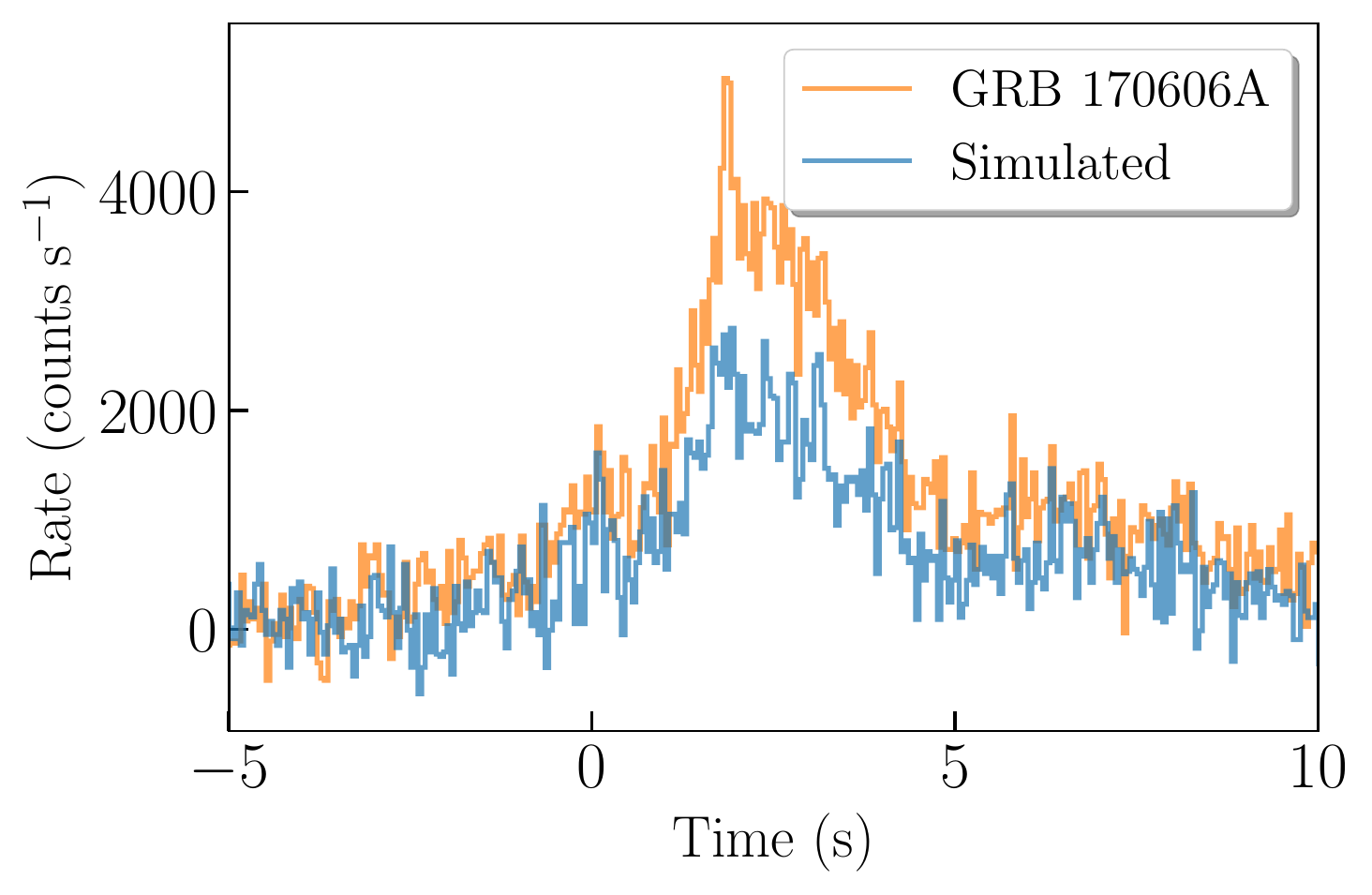}
    \caption{Example of a background subtracted light curve simulated from GRB~170606A, but with a flux level corresponding to that of GRB~160718A (blue), plotted together with the observed light curve of GRB~170606A (orange). Note the similarity with the top left panel of Figure~\ref{fig:160718-170606}. 
    }
    \label{fig:LCsim170606}
\end{figure}

In order to assess the visual appearance of the light curves, it is instructive to consider a specific example of simulated light curves. In Figure~\ref{fig:LCsim170606}, we show the light curve of GRB~170606A together with one of the light curves simulated as described above. From this figure it is clear that the discrepancies between the observed light curves for GRB~160718A and GRB~170606A, seen in the first panel of Figure~\ref{fig:160718-170606}, are not sufficient to rule out a lensing scenario. We thus conclude that CC$_{\text{max}}$ and visual inspection of the light curves suggest that this pair is consistent with a lensing scenario.

\subsection{Impact of observational uncertainties on the spectral analysis}\label{sec:spectralAnalysis}
Although gravitational lensing will not distort spectra, there are other effects that may yield differences in the observed spectra for two lensed GRBs. This includes different observing conditions, such as varying background and angle of incidence to the detectors. Additionally, a change in flux level may affect the spectral fits. Here we investigate to what degree the observed spectral parameters are affected by changes in the flux level, background and response matrix.  Specifically, we consider the candidate pair GRB~160718A-GRB~170606A, which has a significant discrepancy in the posteriors in the second time bin (see Figure~\ref{fig:160718-170606}). 

The relative number of photons in the second time bin for this burst pair is $\sim 0.67$, with GRB~170606A being brighter. 
We start by simulating spectra for this time bin from a set of model parameters drawn from the posterior for a cutoff power law conditioned on the observed data of GRB~170606A in this time bin. We then fit the simulated spectra with a cutoff power law. 
To simulate the lensed spectrum, we draw a new set of parameter values from the original posterior, but adjust the model normalization by a factor $0.67$.
In order to account for the observing conditions of GRB~160718A, we also use the response matrix and background from this burst for the simulations. 
We then sample the posterior for a cutoff power law conditioned on these simulated data as well. These procedures are repeated 100 times.

\begin{figure}
    \centering
    \includegraphics[width=0.45\textwidth]{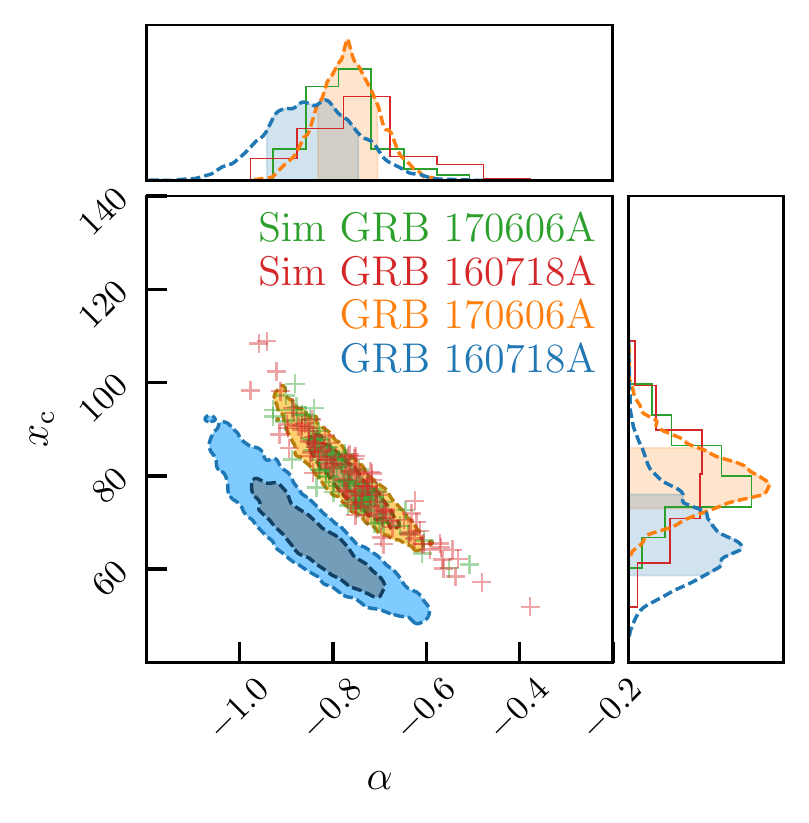}
    \caption{Corner plot showing posteriors of a cutoff power law conditioned on data from bin 2 of GRB~160718A (blue) and GRB~170606A (orange) (cf.  Figure~\ref{fig:160718-170606}). In green we show a collection of 100 posteriors for data simulated from the best-fit model of GRB~170606A. Each point represents the maximum posterior point of a single chain. The red points show the corresponding results for data simulated from the best-fit model of GRB~170606A, but using the flux level, background and response of GRB~160718A.
    }
    \label{fig:fakes}
\end{figure}

Figure~\ref{fig:fakes} shows the results of these simulations. It is clear that the changes in flux, response and background do not resolve the tension between the fits. This is not surprising, since the effect of changing the flux should be an increased spread of the posterior, as can be seen in the figure. The fact that the changes in response and background make little difference is also expected given a well-calibrated instrument and adequate background treatment in the analysis. 
To reconcile the observations of GRB~160718A-GRB~170606A with a common physical origin would require us to invoke some more exotic lensing scenario that can result in the observed differences.

The probability of observing spectral evolution as similar as in GRB~160718A-GRB~170606A among non-lensed GRBs can be estimated from the NL sample. While a time-resolved analysis of the full NL sample is beyond the scope of this work, we have performed a time-resolved analysis of 14 burst pairs from the NL sample. These were selected similarly to the 22 pairs from the L sample in Section~\ref{sec:timeResolvedSpectra}. From these 14 burst pairs we find that GRB~141028A-GRB~190604A have significant overlap in the majority of time bins. This indicates that overlap of the posteriors is not so rare that it constitutes a smoking gun signal for lensing. Thus, as already noted above, we conclude that it is unlikely that GRB~160718A and GRB~170606A are gravitationally lensed GRBs of a common origin.

\section{Results and discussion} \label{sec:Discussion}
Our search for pairs of gravitationally lensed GRBs in a sample of 2712 GRBs observed by {\it Fermi}~GBM did not reveal any convincing candidates. Below, we compare these results with previous studies. We also discuss the implications of the null results, future prospects and similarities within the GRB population. 

All previous searches for macrolensed GRBs have also yielded null results \citep{Veres2009,2011AIPC.1358...17D,Hurley:2019km,Li2014}. These previous works have considered different data sets and used  partly different methods to search for lenses. The most comprehensive search was performed by \cite{Hurley2019}, who analyzed  $\sim 2300$ long GRBs detected by {\it Konus-Wind}. The authors calculated the CC$_{\rm max}$ for all pairs of GRBs and then compared the sky positions and time-averaged spectra for the pairs in the top $0.25\%$ of the CC$_{\rm max}$ distribution. In the study by \cite{Li2014}, the large number of GRBs observed by {\it BATSE} was exploited to search for lenses. The procedure adopted in this case was to select pairs with angular separations $< 4^{\circ}$ and overlapping time-averaged spectral parameters in the {\it BATSE} 5B spectral catalog \citep{Goldstein2013}, and finally compare the light curves of those pairs by eye.

Both \cite{Veres2009} and \cite{2011AIPC.1358...17D} present small, exploratory studies of early data from {\it Fermi}~GBM, where they use the CC to select promising candidates and then compare the time-averaged spectra. \cite{Veres2009} assess the CC through manual inspection, while \cite{2011AIPC.1358...17D} impose constraints on the symmetry of the CC function and its behavior as a function of the temporal resolution of the light curves. It is worth noting that \cite{Veres2009}, \cite{2011AIPC.1358...17D} and \cite{Li2014} all require that the later GRB should be fainter. We have not imposed this constraint since it originates from simple spherically symmetric lens models. Our methods also differ from previous works in that we consider time-resolved spectra and assess the similarities of the most promising pairs using simulations.  This approach offers a powerful way to eliminate candidates.

There are four GRB pairs in our initial sample that have been identified as interesting by previous works. Even though most of these were ultimately rejected in the previous studies, it is worth noting where they landed in our analysis. GRB080730B-GRB090730A and GRB081216A-GRB090429D were identified as interesting candidates in \cite{Veres2009}. In our work they pass the time-averaged spectral and localization cuts, but have low ranks in the CC$_{\text{max}}$ distribution and are not analyzed further. They were ultimately rejected also in \cite{Veres2009}. GRB090516C-GRB090514A, also identified in \cite{Veres2009}, did not pass the time-averaged spectral cut. This burst pair was not rejected by means of spectral information in \cite{Veres2009} due to lack of available detector response matrix at the time. \cite{2011AIPC.1358...17D} point to GRB080804A-GRB081109A as a possible lensed pair, but this pair fails our time-averaged spectral cut and would also be rejected based on the positions available from {\it Swift}.

The discovery of lensed GRBs would be important because the excellent time resolution of GRB detectors offers good prospects for modeling lenses and constraining cosmological parameters.
In addition, the fact that we find no convincing lens candidates could in principle be used to place constraints on the properties of lenses and the GRB population.  However, this is not possible at present since the null result does not rule out the presence of lenses in the sample. Indeed, our analysis has shown that also low values of CC$_{\text{max}}$ are compatible with lensing, and that pairs of GRBs that are known not to be lensed can have very similar spectra and light curves. While the NL sample and the light curve simulations have been very useful for guiding the analysis, they cannot be used to quantify the probability of false positives or negatives. In addition to the simplified nature of the simulations, the main complicating factor is that the L sample is biased to lower SNR. This bias arises because GRBs with low SNR tend to have larger uncertainties on position and spectral parameters, making them more likely to pass the selection criteria. For comparison, the mean SNR for the L sample is about 50 \% lower than that of the NL sample. This bias is not accounted for in the simulated sample, making quantitative comparison to the L sample difficult. 

The main challenges with identifying lensed GRBs from current observational data are the large localization uncertainties and lack of redshift measurements. GRB observations from {\it Swift} are superior to {\it Fermi}~GBM in these respects, but {\it Swift} also has a significantly smaller field of view, which makes the probability of observing a pair of lensed GRBs very low \citep{Li2014}. While the identification of large numbers of lensed GRBs will most likely have to await future missions, the discovery of a lensed pair in the growing samples from current telescopes remains a possibility. As we have shown, it is important to consider both light curves and time-resolved spectra, and to assess properties like CC$_{\text{max}}$ using simulations. 

It is clear that the use of a hard cut on CC$_{\text{max}}$ is the main uncertainty in our search for lenses, with Figure~\ref{fig:CCdistr_L_SL} demonstrating that there is a high probability for lensed pairs to be excluded in this step. By contrast, investigations of our simulated sample suggests that most lensed pairs would pass the initial hard cuts on duration and time-averaged spectra (Section~\ref{sec:samples}), which is expected since these cuts are very conservative. Additionally, lensed pairs that are selected based on a high CC$_{\text{max}}$ are also expected to pass the final time-resolved spectral analysis (Section~\ref{sec:timeResolvedSpectra}). This has been investigated using simulations similar to those described in Section~\ref{sec:spectralAnalysis}. However, we caution that simulations with a realistic lens model (e.g., considering larger variations in flux between the two GRBs in a lensed pair) may give different results. A possible way to improve the use of the CC in future studies is to carry out extensive simulations to assess the CC$_{\text{max}}$ for each pair of GRB, using similar methods as in Section Section~\ref{sec:crossCorrelation}. This may lead to identification of promising candidates that are missed when only selecting based on high values of CC$_{\text{max}}$.

Finally, we note that the search for lensed GRBs has provided information regarding the diversity of GRBs. It is notable that our most promising candidates were single-pulsed GRB, as were most of the 315 GRBs that we investigated based on their high values of CC$_{\text{max}}$. Our results show that some of these GRBs also have very similar spectra. This suggests the existence of a relatively simple physical scenario producing the emission, as well as similarities between the progenitors. Further examination of the properties of these GRBs may help shed light on the nature of GRB progenitors and the origin of the prompt emission.

\section{Summary and conclusions} \label{sec:summary}
We have searched for gravitationally lensed pairs of GRBs, specifically considering the case of macrolensing, in 11 years of \textit{Fermi}~GBM data. The sample consists of about 3.6 million unique pairs. We begin by eliminating burst pairs that are incompatible with a common physical origin based on sky localization, relative duration and time-averaged spectral information available from the \textit{Fermi}~GBM catalog. We then use the CC to investigate the similarity of light curves, and finally analyze the time-resolved spectra of the most promising pairs. We find no convincing cases of gravitationally lensed GRBs. The most similar pairs have single-peaked smooth light curves with relatively few time bins for the spectral analysis. This is best explained by similarities within the GRB population rather than lensing.

We stress that this study does not rule out the existence of gravitationally lensed GRBs in the sample. By simulating light curves, we show that the CC$_{\text{max}}$ distribution compatible with lensed GRBs is broad. This means that a high CC$_{\text{max}}$ alone is not an adequate measure by which to identify lenses. Similarly, a low value of CC$_{\text{max}}$ does not necessarily reject a lensing scenario. We conclude that null-results of studies that rely mainly on the value of CC$_{\text{max}}$ from binned light curves (which includes previously mentioned studies on this topic; \citealt{Hurley:2019km,Li2014,Veres2009,2011AIPC.1358...17D}), cannot be used to make reliable inferences about the lens populations. Constraints on e.g., dark matter distributions derived from such null-results are therefore unreliable. 

To refine the search for lens candidates, it is important to also consider spectral information.  Although time-averaged spectral properties are sufficient to rule out many pairs, we find that a time-resolved spectral analysis is a powerful tool to further eliminate candidates. However, as for CC$_{\text{max}}$, it is clear that a similar spectral evolution on its own does not provide sufficient evidence for lensing. This is evident from the fact that we identify GRB pairs that are known not to be lensed, but which still exhibit similar spectral evolution. 

We conclude that the identification of lensed GRBs requires a comparison of both light curves and time-resolved spectra, and that the significance of the similarities/differences must be assessed with simulations. With these techniques it is possible that convincing lens candidates can be identified in the growing samples of GRBs from current missions. Ultimately, a much larger fraction of well-localized GRBs with redshift measurements is needed to identify lensed GRBs with a high degree of confidence.

\acknowledgments
This work was supported by the Knut \& Alice Wallenberg Foundation.

\vspace{5mm}
\facilities{{\it Fermi} (GBM)}

\software{Scipy \citep{2019arXiv190710121V}
          Astropy \citep{2013A&A...558A..33A}
          3ML \citep{2015arXiv150708343V}
          Seaborn \citep{michael_waskom_2017_883859}
          IPython \citep{PER-GRA:2007}
          }

\appendix

\section{Light curve simulations}\label{appendix:simulations}
In this appendix we describe simulations performed to model lensed GRB light curves. The purpose is to assess the CC$_{\text{max}}$ distribution and typical changes in light curves in a simplified lensing scenario. We do not consider the redshift distribution of GRBs, any specific lens model or the smearing of light curves expected due to microlensing.  We instead perform the simulations based on randomly chosen light curves from the GRB sample, with simple assumptions on how signal and background vary due to the observing conditions and lensing. The CC between the simulated and original light curves are  calculated to give the CC$_{\text{max}}$ distribution.

We base the simulations on observed light curves with $0.05$~s time bins. Each time bin in such a light curve has a signal corresponding to a Poisson rate $\lambda_i$ and a background count rate $b_i$, where $i$ indicates the time bin. We simulate the signal and background separately and then subtract the background, as described in Section~\ref{sec:lightcurves}. 
%%%%%%%
When simulating the signal photons, we must address the negative count rates present in some bins as a result of the background subtraction. Since rates will be negative only in bins that are essentially all background, we simply simulate no signal in these bins. Additionally, we only simulate signal counts inside $T_{90}$ in order to avoid biasing the results due to this effect. We simulate photon arrival times in each time bin using a rate $\lambda_{i,\text{sim}} = c_\text{s} \lambda_{i}$, where $\log c_\text{s} \sim \mathcal{U}(0.7,0.7^{-1})$. This interval is not motivated by a specific lens model, but is chosen as a conservative model for plausible changes in flux levels for lensed GRBs. The factor $c_\text{s}$ is kept constant for each realization of a simulated light curve. 

For the background we simulate the arrival times of photons using a rate $b_{i,\text{sim}} = c_\text{b} b_i$ in each time bin, where $c_\text{b} \sim \mathcal{N}(\mu = 1,\sigma = 0.1)$. 
Again, $c_\text{b}$ is kept constant for each realization of a simulated light curve. We concatenate the list of arrival times and then bin them in time bins of $0.5$ and $0.05$~s. However, to acknowledge some variability in the trigger time caused by Poisson fluctuations and imperfections in the trigger algorithm, we create the bins starting from an offset drawn from $\mathcal{U}(0,\Delta)$, where $\Delta$ is the bin width. Without this shift we would significantly inflate the CC$_{\text{max}}$ calculated from the simulations relative to a corresponding CC$_{\text{max}}$ in the L sample. 

\begin{figure*}{}
    \centering
        \includegraphics[width=0.42\textwidth]{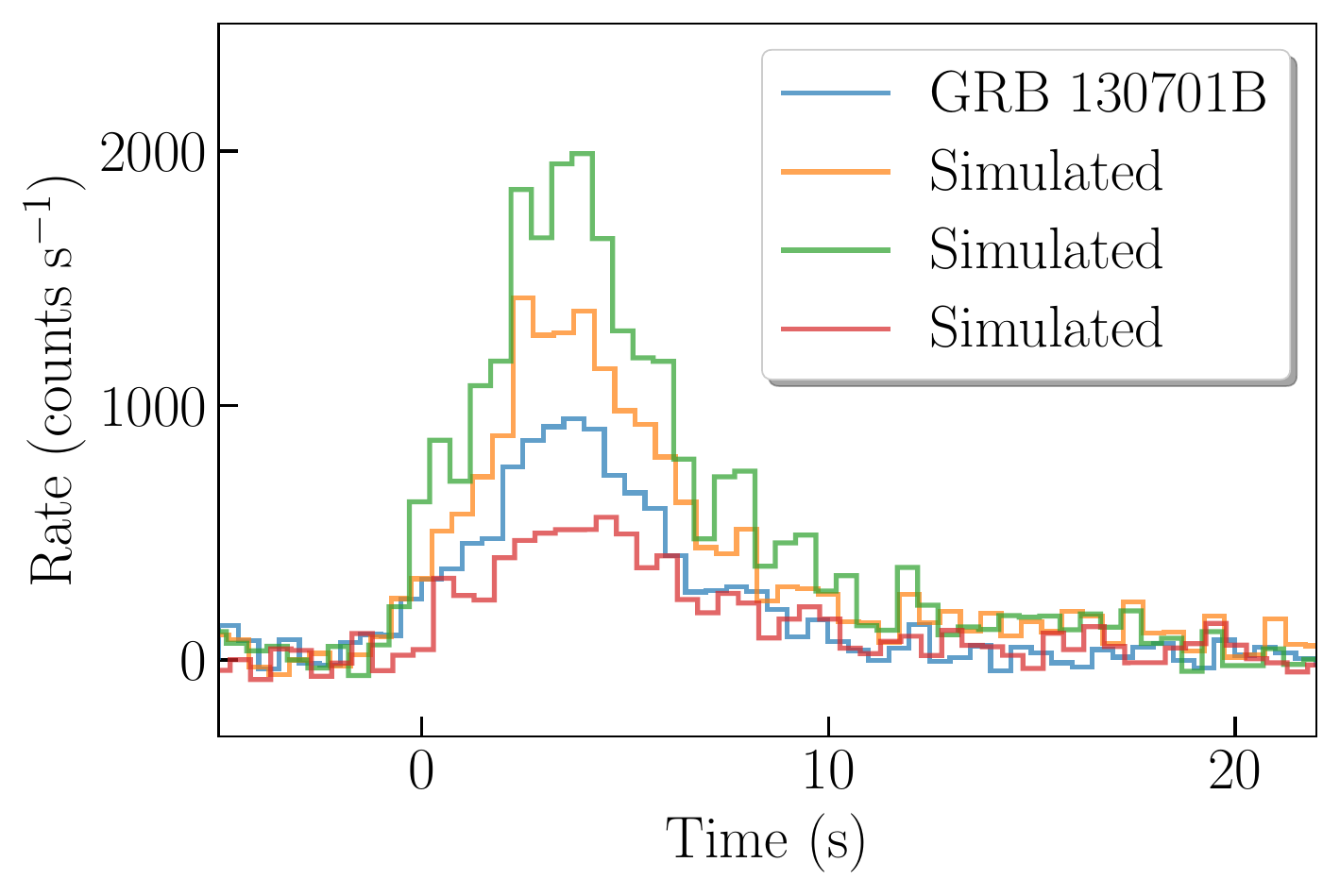}
        \includegraphics[width=0.42\textwidth]{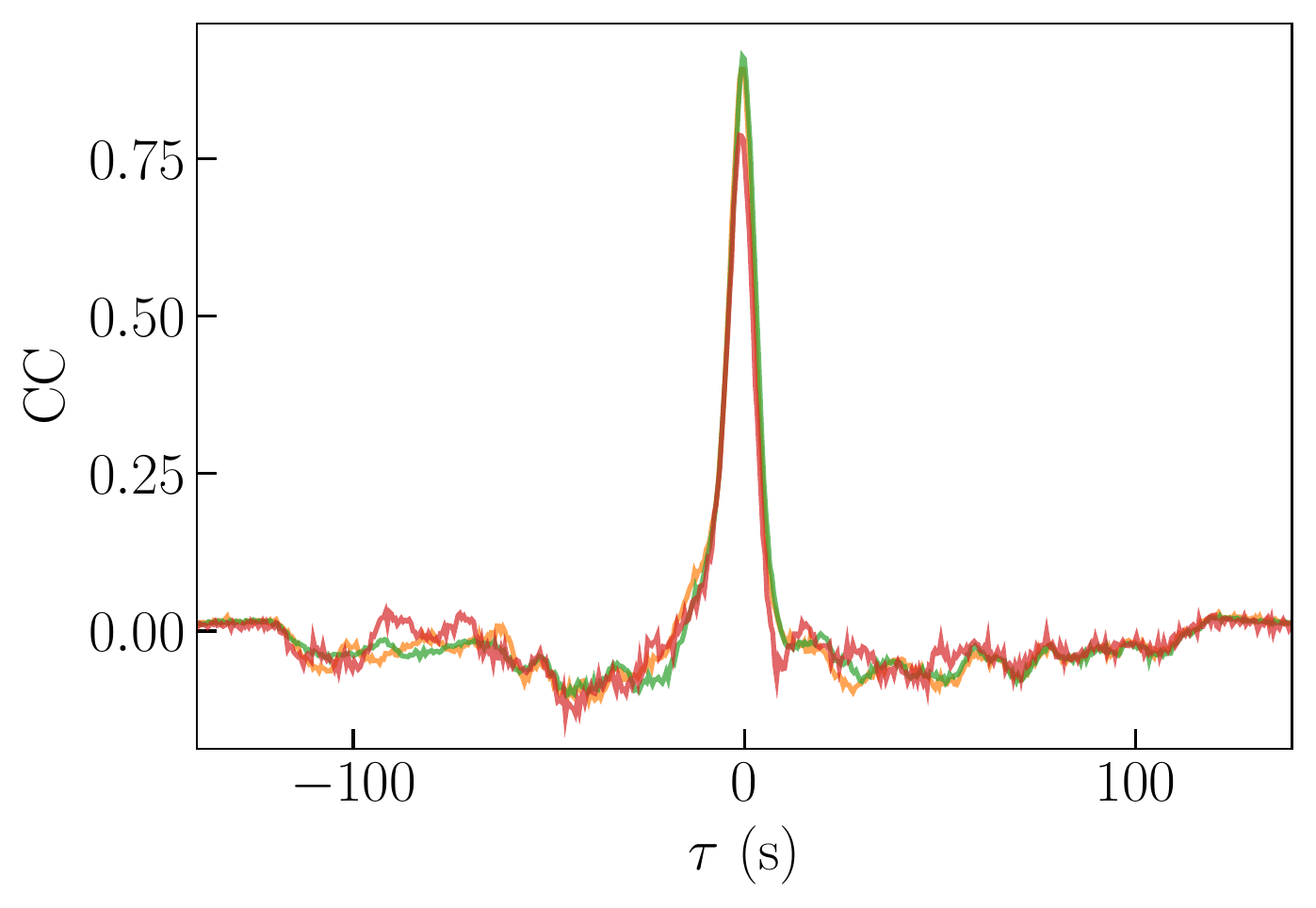}
        \includegraphics[width=0.42\textwidth]{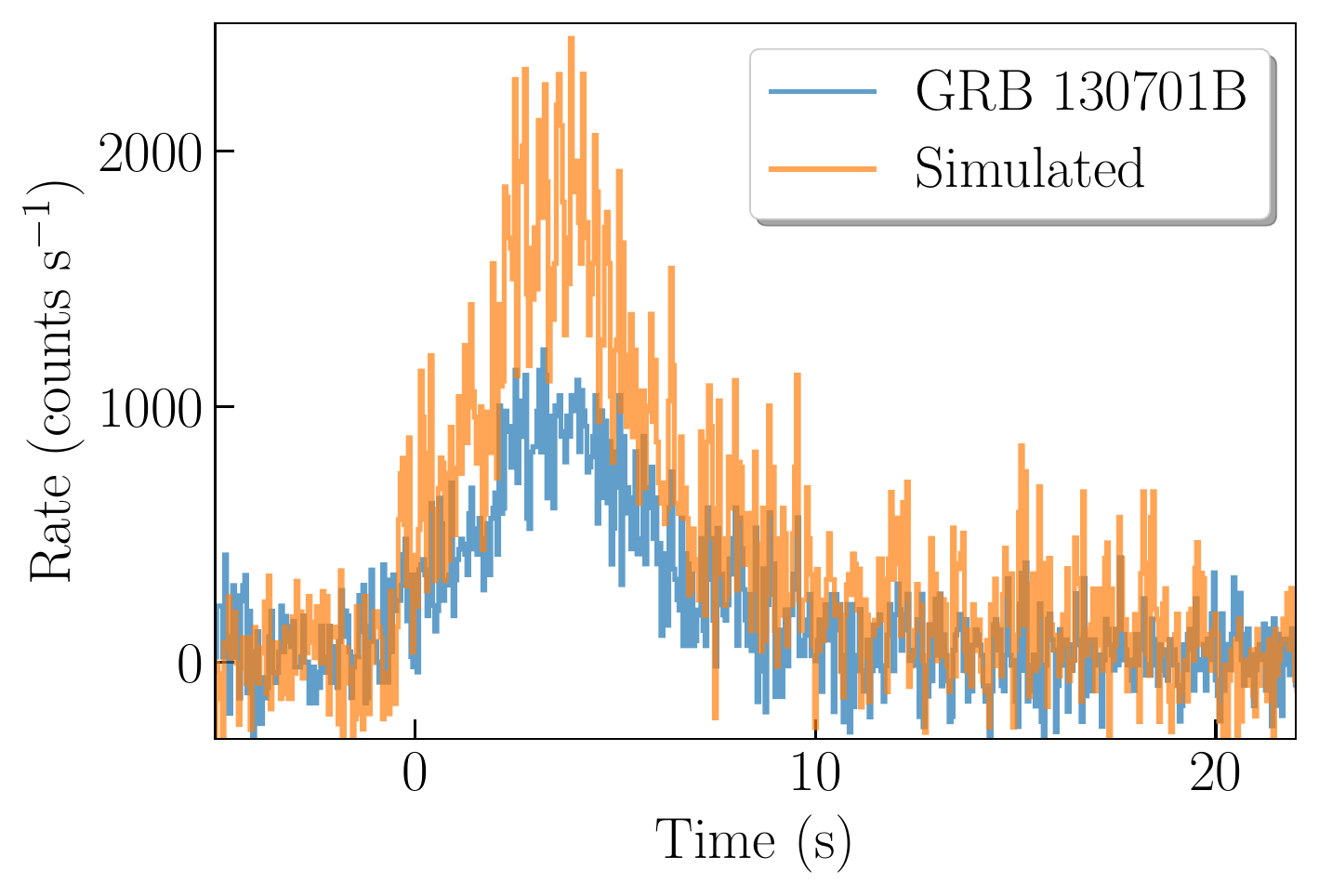}
        \includegraphics[width=0.42\textwidth]{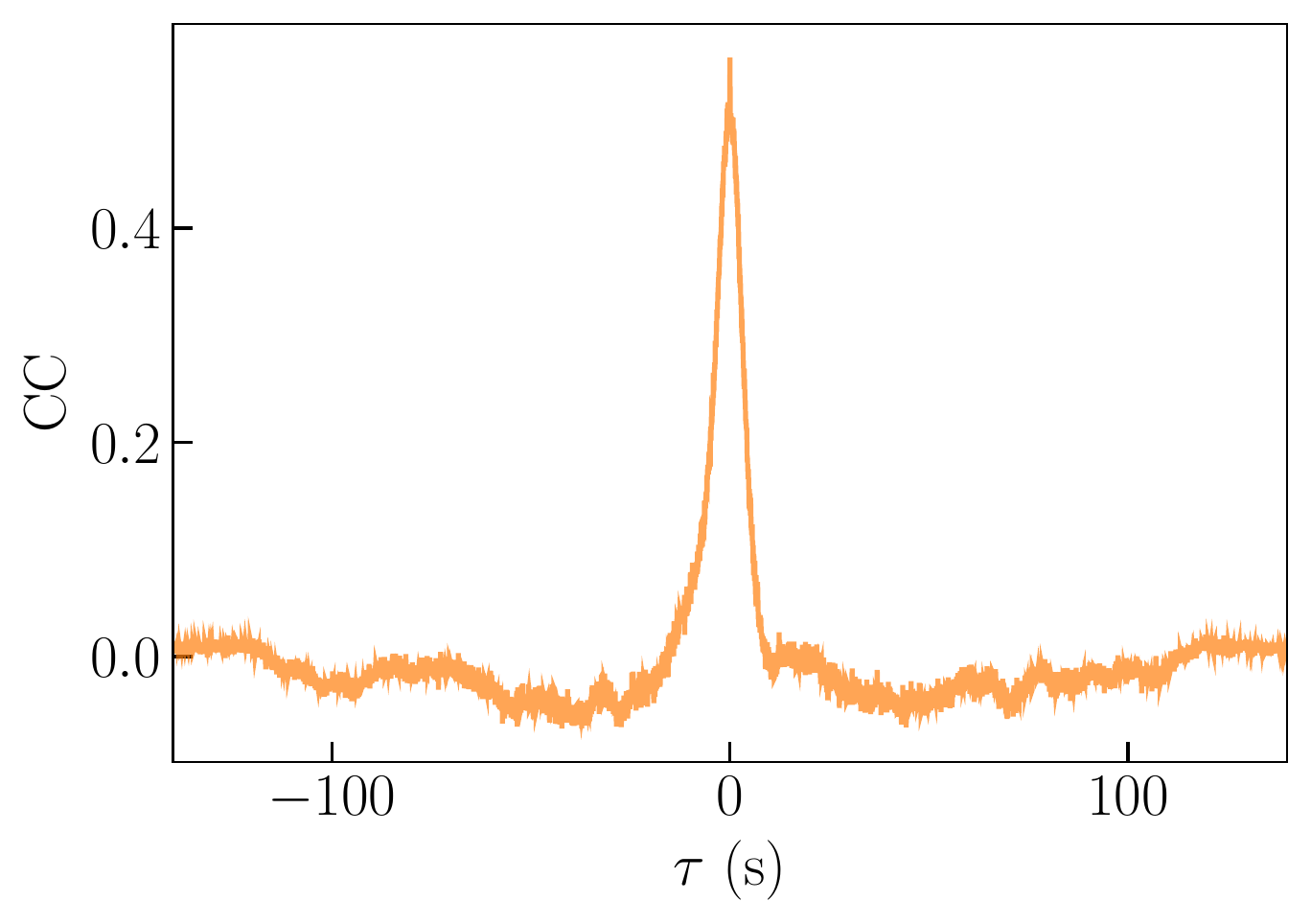}
    \caption{Examples of light curves simulated from GRB~130701B and their corresponding CCs when compared to the original light curve of GRB~130701B. The blue line shows the original light curve, while the color of the CC indicates which of the simulated light curves it represents. The top and bottom rows show results for 0.5~s and 0.05~s time bins, respectively. We show only one simulated light curve for the $0.05$~s time bins for visual clarity.}
    \label{fig:simulatedLCexamples}
\end{figure*}

We simulate 10 light curves for each of 550 randomly chosen observed light curves and calculate CC$_{\text{max}}$ for the simulated light curves with respect to the original ones. By considering 550 randomly chosen light curves, we cover much of the observed light curve morphology. The resulting distributions of 5500 CC$_{\text{max}}$ for the two time bins are presented in Figure~\ref{fig:CCdistr_L_SL}. To illustrate typical changes in light curves, we also show examples of simulated light curves and the corresponding CCs for GRB~130701B in Figure~\ref{fig:simulatedLCexamples}. It is clear that the small changes introduced by the change in flux and trigger time, as well as pure Poisson fluctuations, can change the look of GRB light curves significantly. This provides further motivation for not using visual inspection of  light curves to gauge the probability of lensing.

\bibliography{bibFile.bib}
\bibliographystyle{aasjournal}

\end{document}